\documentclass[aps,pre,showpacs, amsmath,
  amssymb,reprint,groupedaddress]{revtex4-1}
\renewcommand\vec[1]{\overrightarrow{#1}}

\usepackage{color}
\usepackage{graphicx}
\usepackage{comment}
\usepackage{mathtools}
\begin{document}
\title{Entropy balance and Information processing in bipartite and non-bipartite composite systems}
\author{Richard E. Spinney, Joseph T. Lizier and Mikhail Prokopenko}
\affiliation{Complex Systems Research Group, School of Civil Engineering, Faculty of Engineering and IT,
The University of Sydney, Sydney, New South Wales, Australia, 2006. 
} 

\date{\today}

\begin{abstract}
Information dynamics is an emerging description of information processing in complex systems which describes systems in terms of intrinsic computation, identifying computational primitives of information storage and transfer. In this paper we make a formal analogy between information dynamics and stochastic thermodynamics which describes the thermal behaviour of small irreversible systems.  As stochastic dynamics is increasingly being utilized to quantify the thermodynamics associated with the processing of information we suggest such an analogy is instructive, highlighting that existing thermodynamic quantities can be described solely in terms of extant information theoretic measures related to information processing. In this contribution we construct irreversibility measures in terms of these quantities and relate them to the physical entropy productions that characterise the behaviour of single and composite systems in stochastic thermodynamics illustrating them with simple examples. Moreover, we can apply such a formalism to systems which do not have a bipartite structure. In particular we demonstrate that, given suitable non-bipartite processes, the heat flow in a subsystem can still be identified and one requires the present formalism to recover generalizations of the second law. In these systems residual irreversibility is associated with neither subsystem and this must be included in the these generalised second laws. This opens up the possibility of describing all physical systems in terms of computation allowing us to propose a framework for discussing the reversibility of systems traditionally out of scope of stochastic thermodynamics.
\end{abstract}
\maketitle
\section{Introduction}
In the early 1990s Evans and co-workers  \cite{evans_probability_1993} first measured the probability of ``second law violations'' that occur in small systems. This new found ability to quantify thermodynamic quantities at scales where uncertainties dominate  opened up a new frontier in modern thermodynamics leading, notably, to the now famous work relations of Crooks and Jarzynski \cite{crooks_entropy_1999,crooks_pathensemble_2000,jarzynski_nonequilibrium_1997,jarzynski_equilibrium_1997}. Later a unifying framework for general stochastic systems \cite{seifert_entropy_2005} has formed the basis for a now indispensable tool for the study of small noisy systems, named \emph{stochastic thermodynamics} \cite{seifert_stochastic_2008,seifert_stochastic_2012}. More recently, however, such formalisms have implicated information theoretic measures introduced in contexts such as the exploitation of pure order \cite{barato_stochastic_2014,mandal_work_2012,deffner_information_2013,chapman_how_2015,mcgrath_biochemical_2017}, or more generally where feedback and/or measurement, is occurring \cite{horowitz_nonequilibrium_2010,horowitz_thermodynamics_2014,horowitz_secondlawlike_2014,parrondo_thermodynamics_2015,esposito_stochastic_2012,sagawa_second_2008,sagawa_generalized_2010,sagawa_nonequilibrium_2012,toyabe_experimental_2010,still_thermodynamics_2012,strasberg_thermodynamics_2013,abreu_thermodynamics_2012} providing powerful relationships between physical quantities and information. A concerted effort is now being levelled at utilizing such a connection to understand the thermodynamics of computation \cite{sagawa_minimal_2009,kempes_thermodynamic_2017,wolpert_extending_2015,esposito_second_2011-1,prokopenko_information_2015,prokopenko_transfer_2014}.
\par

Quite separately, however, there is an alternative viewpoint of physical (or otherwise) processes, which posits that they support \emph{intrinsic computation} \cite{crutchfield_calculi_1994,feldman_organization_2008}. In this view subsequent states of a dynamical process are considered to have been computed from earlier ones. In practice, however, the confidence  in the precise values of such subsequent states is never perfect (arising through inherent coarse-graining, imprecise initial conditions or, in abstract, pure noise). Given this unavoidable reality, the central approach becomes the quantification of these uncertainties in a non-parametric manner through the introduction of information theory. This in turn provides a natural viewpoint for the characterisation of the information processing supported by the dynamical process. In this view, the exact structure of physical dynamics give way to the phenomenological predictive capacities/probabilities derived from them, from which, in turn, the structure of the information processing performed by the system can be characterized.
\paragraph*{Background}
 Framing such a discussion is the concept of generalised computation and a challenge put forward by Langton \cite{Lan91,langton_computation_1990} 
in the early 1990s: how can emergence of computation and, in particular, the emergence of complex spatiotemporal dynamics capable of universal computation, 
be explained in a dynamic setting.  In doing so, he suggested to dissect computation into three primitive functions: the \emph{transmission}, \emph{storage} and \emph{modification of information}. Over the last decade, this decomposition has been formalized through an emerging field called information dynamics \cite{lizier_local_2013,lizier_framework_2014,bossomaier_introduction_2016,lizier_local_2008,lizier_multivariate_2010,lizier_information_2010,lizier_local_2012, williams_information_2010,walker_informational_2016,faes_conditional_2014,dasgupta_information_2013} proving itself to be useful in revealing or explaining  generic mechanisms underlying dynamics of distributed information processing. 
\par
A common feature in several studies which used the primitive information-processing functions in tracing information dynamics is the distributed but coherent nature of the underlying computation~\cite{liz10d}. For example, the coherent behavior  given by swarms of animals that self-organize in complex large-scale spatial patterns is often explained by ``collective memory'' \cite{COUZIN2002} and ``collective communications'' expressed via wave-like cascades of information transfer~\cite{couz06}. Interestingly, the information cascades have been conjectured to correspond to long range communications that either dynamically reorganize the swarm reducing the ``fragility of mass behaviour''~\cite{BHW_jpe1992} or propagate incorrect decisions~\cite{Galef_AB2001}. 
\par
It has been conjectured that information plays a dynamic role in such a self-organization~\cite{sumpter2008}, and more specifically, that distributed information-processing enables the groups to collectively perform computation~\cite{bonabeau1999swarm, couzin2009collective, albantakis2014evolution}. 
These conjectures have been formalized in the context of information dynamics \cite{Wang12}, verifying the hypothesis that the collective memory within a swarm can be captured by \emph{active information storage}, while the information cascades are captured by \emph{conditional transfer entropy}~\cite{liz08a,liz10e}. This has been further verified for real biological groups, such as swarms of soldier crabs~\cite{tomaru2016}, schools of zebrafish responding to a robotic animal replica~\cite{butail2014}, pairs of bats~\cite{orange2015transfer}, rummy-nose tetras (\emph{Hemigrammus rhodostomus}) fish schools~\cite{Crosato2018}, and so on.  
\par
A more abstract set of examples is given by dynamics of random Boolean networks --- a canonical model for Gene Regulatory Networks (GRNs)~\cite{kauf93}. When the average connectivity or activity level is varied, the dynamics undergo a phase transition between order and chaos, and it has been shown that the information storage and information transfer are maximized on either side of the critical point: for fully-random topologies the information storage dominates the ordered phase of dynamics and the information transfer is the primary information-processing primitive at the chaotic phase~\cite{liz08c}. When the underlying topology changes as well, specifically when it undergoes an order -- small-world -- randomness transition, the interplay between information storage and transfer attains a balance so that the network near the small-world state retains a sufficient amount of storage capability while being able to transfer information at a high rate~\cite{liz11b}. 
\par
Information dynamics methods have also been successfully used for online machine learning in robotic and artificial life scenarios~\cite{Ay2008-PI,der2012playful,Ma2007,Martius2010,Der2012}, again allowing the designers to guide the automated learning process along specific information-processing primitives and fine-tune the computational balances, 
in addition to providing novel insights across a broad range of fields including in: canonical complex systems such as cellular automata \cite{lizier_local_2013}, interpretation of dynamics in \cite{williams_information_2010} and improved algorithms for machine learning \cite{dasgupta_information_2013}, characterizing information processing signatures in biological signalling networks \cite{walker_informational_2016},  
non-linear time-series forecasting \cite{gar16a}, and in computational neuroscience applications in identifying neural information flows from brain imaging data \cite{faes_conditional_2014,timme_multiplex_2014,ito_extending_2011}, inferring effective network structure \cite{faes15b,wibral_transfer_2011,vicente_transfer_2011}, providing evidence for the predictive coding hypothesis \cite{brodski_2017} and identifying differences in predictive information in autism spectrum disorder subjects and controls \cite{gomez_2014}.
\par
Despite these developments, the character of physical entropic balances between the information-processing primitives engaged in a distributed computation has not been fully elucidated. 
And indeed, until now the separate perspectives of stochastic thermodynamics and information dynamics have not been comprehensively understood in terms of each other.

{As such our study has the following motivation --- how might the computational primitives described in information dynamics manifest themselves in thermodynamics? ---  and in what way might physical bounds or constraints upon the system be expressed solely in terms of  the computational primitives that the system supports? And do such quantities point to a dynamic generalization of the current thermodynamic bounds purely in terms of information theoretic characterisations of the dynamics?
\paragraph*{Outline} In this paper, in order to answer the above questions, we explicitly introduce the notion of irreversibility into the language of distributed computation and in doing so make a formal connection between information dynamics and stochastic thermodynamics, specifically in terms of stored and transferred information and the bounds they place on each other. 
\par
Through the definition of an appropriate \emph{time reversed computation}, we discuss the storage and transfer of information of time reversed behaviour and consequently identify irreversibility measures associated with stored and transferred information. Importantly, for physical systems which permit identification of an unambiguous entropy production, this total entropy production is identically equal to the sum of irreversibilities associated with storage and transfer of information. This allows us to identify contributions to the physical entropy production solely in terms of computational primitives.
\par
We illustrate the behaviour of such entropy production contributions in a variety of increasingly broad situations. First we consider the canonical situation in stochastic thermodynamics, namely that of a system controlled by an external protocol, illustrating that the entropy production associated with information transfer to the protocol is intimately related to the irreversibility in the dynamics of that protocol, demonstrating different bounds on the contributions for reversible and irreversible protocol dynamics. We then apply the framework to composite systems, considering first bipartite systems. We show that our framework identifies the physical entropy contributions which allow the second law to be locally broken for individual subsystems, for instance in the context of feedback, in contrast to previous approaches and results which aim to bound such terms with information theoretic measures. 
In considering such composite systems, we introduce a principle of conserved predictive capacity, which constrains the extent to which computational primitives of the total system can be associated with the individual subsystems. We find that such a principle always leads to the existence of an interaction term understood to be a predictive capacity not uniquely attributable to any individual subsystem. Further this interaction term is essential for constructing generalizations of the second law in terms of stored and transferred information. For bipartite systems this interaction term reduces to the change in mutual information recovering previous results where such terms are introduced \emph{ad-hoc} in order to construct the appropriate bounds.
\par
Moreover, because our formalism identifies contributions to the entropy production using physically agnostic information theoretic measures, the resultant relations are valid where previous ones are not. Specifically, they apply non bipartite systems where we can still associate computational irreversibility despite unambiguous heat flows not necessarily being well defined. Importantly, in these systems the relations from our formalism that hold for bipartite systems remain \emph{unchanged}, naturally accounting for intrinsic correlation in the dynamics of the subsystems. We emphasize that systems which are physically meaningful are not restricted to the bipartite case, yet previously no theory of how entropy production should be associated with the behaviour of individual subsystems has been provided. As such we posit that such a framework is a natural approach to the phenomena of irreversibility, applicable to all distributed computing systems, from which known bounds in stochastic thermodynamics emerge where the relevant physical assumptions hold. 
\par
We illustrate our results on two simple models. The first is a simple bipartite model allowing illustration of both entropy associated with storage and transfer of information. The second is a non-bipartite linear model of a feedback controller with system and controller driven by correlated noise. Until now we are aware of no formalism for describing this set up. We emphasize that physical heats are identifiable in such a system, that previous generalizations of the second law with information theoretic terms fail in this context, but the bounds from the present formalism do hold for any degree of correlation in the dynamics.

\section{Intrinsic computation and information dynamics}
\label{infodyn}
The outlook that intrinsic computation may be decomposed into distinct primitives naturally frames the discussion of computation in terms of identifying precise measures of information processing which information dynamics \cite{lizier_local_2013,lizier_framework_2014,bossomaier_introduction_2016,lizier_local_2008,lizier_multivariate_2010,lizier_information_2010,lizier_local_2012, williams_information_2010,walker_informational_2016,faes_conditional_2014,dasgupta_information_2013} aims to quantify.
\par
Originally formulated in the context of distributed computing systems, possessing a characteristic update interval, in discrete time, the central object in information dynamics is the \emph{predictive capacity} which quantifies the transformation of uncertainty in the `output', some next state in a given time series, when all previous `input' states, all current and previous time series data, are known. 
This predictive capacity, measured in bits or nats, is characterized by a mutual information, defined between two variables $A$ and $B$ taking values $a\in\mathcal{A}$ and $b\in\mathcal{B}$ respectively, as
\begin{align}
I_M(A;B) &= \sum_{\substack{a\in\mathcal{A}\\b\in\mathcal{B}}}p(A=a,B=b) \ln\frac{ p(A=a,B=b)}{p(A=a)p(B=b)}\nonumber\\
&=\left\langle \ln\frac{ p(A=a,B=b)}{p(A=a)p(B=b)}\right\rangle,
\end{align} 
such that $\langle\ldots\rangle$ indicates an ensemble expectation. Specifically, the predictive capacity is the mutual information between the complete past of the universe and the next computed state of the variable in question. We characterise the evolution of the variable in question, $X$, taking values $x\in\mathcal{X}$, as a time series $X_{0:i}\equiv\{X_0,X_1,\ldots,X_i\}$, taking values $x_{0:i}\equiv\{x_0,x_1,\ldots,x_i\}$ in a space $\mathcal{X}_{0:i}=\mathcal{X}^{\otimes(i+1)}$, such that $x_i=x(t=i)$. Similarly, the rest of the universe is captured by an extraneous variable $Y$  taking values in $y\in\mathcal{Y}$ which equally is characterized as a time series $Y_{0:i}\equiv\{Y_0,Y_1,\ldots,Y_i\}$, taking values $y_{0:i}\equiv\{y_0,y_1,\ldots,y_i\}$ in a space $\mathcal{Y}_{0:i}=\mathcal{Y}^{\otimes (i+1)}$. As such we consider the predictive capacity, $C_X^{\{i,i+1\}}$, from time $t=i$ to $t={i+1}$, for a process started at time $t=0$ as
\begin{align}
C^{\{i,i+1\}}_X&=I_M(X_{i+1};X_{0:i}Y_{0:i})\nonumber\\
&=H(X_{i+1})-H(X_{i+1}|X_{0:i},Y_{0:i}).
\end{align}
The superscripts indicate, explicitly, that this prediction occurs over the interval $\{i,i+1\}$. In this sense we understand that it characterises the total reduction in uncertainty that can be achieved about the `computed state', $X_{i+1}=x_{i+1}$, given the inputs, namely the previous values of the time series, $\{X_{0:i},Y_{0:i}\}=\{x_{0:i},y_{0:i}\}$. The information theoretic quantities are then built from ensemble probabilities $p(X_i=x_i)$ and transition probabilities $p(X_{i+1}=x_{i+1}|X_{0:i}=x_{0:i})$ such that
 \begin{align}
 &H(X_{i+1})=-\sum_{x_{i+1}\in\mathcal{X}_{i+1}}p(X_{i{+}1}=x_{i{+}1})\ln p(X_{i{+}1}=x_{i{+}1})\nonumber\\
  &H(X_{i+1}|X_{0:i},Y_{0:i})\nonumber\\
  &\quad=-\sum_{\substack{x_{0:i+1}\in\mathcal{X}^{\otimes(i+2)}\\ y_{0:i}\in\mathcal{Y}^{\otimes(i{+}1)}}}p(X_{0:i{+}1}{=}x_{0:i{+}1},Y_{0:i}=y_{0:i})\nonumber\\
  &\qquad\qquad\times\ln p(X_{i+1}=x_{i+1}|X_{0;i}=x_{0:i},Y_{0:i}=y_{0:i})
 \end{align}
 etc. where here and throughout we will assume time homogeneity such that any and all time variation in the dynamics/transition probabilities experienced by $X$ is parametrized through $Y$.
 \par
 The insight from information dynamics is to divide this total predictive capacity into computationally relevant quantities in the spirit of Langton's conception of generalised computation. The pertinent division is that which separates the predictive capacity due to the process $X$ and the predictive capacity over and above that of $X$ due to $Y$. This is written
\begin{align}
C^{\{i,i+1\}}_X&=H(X_{i+1})-H(X_{i+1}|X_{0:i})\nonumber\\
&\quad+H(X_{i+1}|X_{0:i})-H(x_{i+1}|X_{0:i},Y_{0:i})\nonumber\\
&=I_M(X_{i+1};X_{0:i})+I_M(X_{i+1};Y_{0:i}|X_{0:i})\nonumber\\
&=A^{\{i,i+1\}}_X+T^{\{i,i+1\}}_{Y\to X},
\end{align}
where $I_M(A;B|C)=I_M(A;B,C)-I_M(A;C)$ is a conditional mutual information between $A$ and $B$ given $C$. Here $A^{\{i,i+1\}}_X$ is known as the \emph{active information storage} \cite{lizier_local_2012} and $T^{\{i,i+1\}}_{Y\to X}$ as the \emph{transfer entropy} \cite{schreiber_measuring_2000}, a well known non-parametric measure of predictability used in many areas of science, in this instance from $Y$ to $X$. We denote this division of the predictive capacity, the \emph{computational signature} of the process (such a division has been explored previously, see for example \cite{lizier_coherent_2012}). The active information storage is interpreted as the stored information in $X$ which is utilized in making a transition, whilst the transfer entropy is interpreted as the transferred information from $Y$ utilized in making a transition in the context of the full history of $X$.
 Importantly, since both $A_X$ and $T_{Y\to X}$ can be written as (conditional) mutual informations they are thus rigorously non-negative.\\
\\
Recent developments have stressed that these quantities should be understood as expectation values of suitable specific, pointwise or \emph{local}, quantities \cite{robert_fano_transmission_1961,lizier_local_2012}. These are known as the \emph{local active information storage} $a_x$ and \emph{local transfer entropy} $t_{y\to x}$ where we write
\begin{align}
A^{\{i,i+1\}}_X&=\langle a^{\{i,i+1\}}_x\rangle\nonumber\\
T^{\{i,i+1\}}_{Y\to X}&=\langle t^{\{i,i+1\}}_{y\to x}\rangle,
\end{align}
with
\begin{align}
a^{\{i,i+1\}}_x&=\ln\frac{p(X_{i+1}=x_{i+1}|X_{0:i}=x_{0:i})}{p(X_{i+1}=x_{i+1})}\nonumber\\
t^{\{i,i+1\}}_{y\to x}&=\ln\frac{p(X_{i+1}=x_{i+1}|X_{0:i}=x_{0:i},Y_{0:i}=y_{0:i})}{p(X_{i+1}=x_{i+1}|X_{0:i}=x_{0:i})}
\end{align}
which together comprise the local predictive capacity
\begin{align}
c^{\{i,i+1\}}_x&=a^{\{i,i+1\}}_x+t^{\{i,i+1\}}_{y\to x}\nonumber\\
&=\ln\frac{p(X_{i+1}=x_{i+1}|X_{0:i}=x_{0:i},Y_{0:i}=y_{0:i})}{p(X_{i+1}=x_{i+1})}
\end{align}
with $C^{\{i,i+1\}}_X=\langle c^{\{i,i+1\}}_x\rangle$. In purely information theoretic terms, such local quantities may be interpreted as the differences in code lengths between competing models of the observed behaviour in $X_{i+1}$. It is important to note that such local values have no bound on their sign and thus may be negative. Such an approach allows significance to placed on \emph{single realisations} of a process, allowing fine characterisation of spatial temporal features, such as the identification of dynamics that are informative, but especially those which are \emph{misinformative}, characterized by negative local values. Detailed understanding of these \emph{individual} realisations using such local quantities has yielded important identification and insights into distributed computing behaviour. This ability to attribute storage and transfer of information to individual realisations is crucial to the following developments.
\section{Time reversed computations}
\label{revcomp}
Modern thermodynamics makes explicit links between the dynamic irreversibility of a physical process and the total entropy production of the universe. Indeed, this `total entropy production', comprising the change in internal uncertainty and the heat dissipated to an environment, can be, under the right circumstances, considered equal to the log-ratio of the probability of observing the `forward' behaviour relative to the probability of observing the `reverse' behaviour, having started from an ensemble characterized by the distribution at the end of the forward process. This entails, if not the construction, the consideration of, a suitable `reverse process' with which the normal, forward dynamics are contrasted.
\par
We wish to give an alternative account of irreversibility in \emph{computational} terms, understood in the sense of information dynamics. To do so we need to attribute computational quantities to the time reverse behaviour of the system. Accordingly, and in contrast to the reverse process considered in stochastic thermodynamics, we define a \emph{time reversed computation} through an analogous \emph{time reversed predictive capacity}. Such a object is defined axiomatically, on a local, pointwise, scale, as the predictive capacity with all constitute elements time reversed relative to the transition being considered in the context of the whole process from  $t_0$ to $t_n$, viz.
\begin{align}
c^{\dagger,\{n{-}(i{+}1),n{-}i\}}_x&=\ln\frac{p^\dagger(X_{i+1}=x^\dagger_{i+1}|X_{0:i}=x^\dagger_{0:i},Y_{0:i}=y^\dagger_{0:i})}{p^\dagger(X_{i+1}=x^\dagger_{i+1})},
\end{align}
noting that this has been defined for the transition from time $n-i-1\to n-i$. Here we have introduced
\begin{itemize}
\item The time reversed transform denoted by the $\dagger$ symbol which time reverses and reflects sequences on the interval $[0,n]$. This leads to the time reversed path $x^\dagger_{0:n}$, which, in the absence of odd parity variables amounts to the reverse sequence of $x_{0:n}$, i.e. $x^\dagger_{0:n}=x_{n:0}=\{x_n,\ldots,x_0\}$, such that $x^\dagger_i=x_{n-i}$. 
\item The time reversed ensemble probability, appearing in the denominator, defined as  $p^\dagger(X_i=x^\dagger_i)=p(X^\dagger_i=x^\dagger_i)=p(X_{n-i}=x_{n-i})$.
\item The \emph{time reversed dynamics} \cite{crooks_nonequilibrium_1998,seifert_entropy_2005,spinney_fluctuation_2013}, appearing in the numerator, which in contrast  accounts for any differences that occur under time reversal in the generator of the process. This can arise due to dependence on odd quantities with respect to time reversal (e.g. magnetic fields/torques etc.) or differences in evolution related to intervention in the system, e.g. in extraneous variables $Y$. Unless otherwise stated we will consider $p^\dagger=p$ relating to autonomous, time invariant, behaviour.
\end{itemize}
We emphasize that the probability in the numerator is of the reverse trajectory as if it were evolving forward in time under dynamics denoted $p^\dagger$.  In this sense it remains a \emph{prediction} of a hypothetical future evolution of the system.  We emphasize that $p^\dagger$ is \emph{not} a probability associated with \emph{retrodiction} i.e. \emph{not} $p^\dagger(X^\dagger_{i+1}=x^\dagger_{i+1}|X_{0:i}^\dagger=x^\dagger_{0:i},Y^\dagger_{0:i}=y_{0:i})$.
\par
For brevity we subsequently drop the explicit reference to the variables and write the probabilities solely in terms of the realized values such that we write
\begin{align}
c_{x}^{\{i,i+1\}}&=\ln\frac{p(x_{i+1}|x_{0:i},y_{0:i})}{p(x_{i+1})}\nonumber\\
c_{x}^{\dagger,\{n{-}i{-}1,n{-}i\}}&=\ln\frac{p^\dagger(x^\dagger_{i+1}|x^\dagger_{0:i},y^\dagger_{0:i})}{p^\dagger(x^\dagger_{i+1})}\nonumber\\
&=\ln\frac{p^\dagger(x_{n{-}i{-}1}|x_{n:n{-}i},y_{n:n{-}i})}{p(x_{n{-}i{-}1})}
\end{align}
and so, for the transition $i\to i+1$,
\begin{align}
c_x^{\dagger,\{i,i+1\}}&=\ln\frac{p^\dagger(x_{i}|x_{n:i+1},y_{n:i+1})}{p(x_{i})}
\end{align}
where we may understand $p(x_i)\equiv p(X_i=x_i)$, $p(x_{i+1}|x_i)\equiv p(X_{i+1}=x_{i+1}|X_i=x_i)$, $p^\dagger(x_i|x_{i+1})\equiv p^\dagger(X_{n-i}=x_i|X_{n-i-1}=x_{i+1})$ and so on. We also note that by time homogeneity we may also write $p^\dagger(X_{n-i}=x_i|X_{n-i-1}=x_{i+1})=p^\dagger(X_{i+1}=x_i|X_{i}=x_{i+1})$. 
Some of these quantities, in the context of continuous time, are illustrated in Fig.~(\ref{fig0}). Here we observe two paths realisations $x_{[t_0,\tau]}\equiv \{x(t), t\in[t_0,\tau]\}$ and $y_{[t_0,\tau]}\equiv \{y(t), t\in[t_0,\tau]\}$ of $X$ and $Y$, defined between the time origin $t_0$ ($t=0$ in discrete time) and time horizon $\tau$ ($t=n$ in discrete time) illustrated in full in the first subplot. The second subplot concerns the prediction associated with the intrinsic computation at a given time $t$ which can be understood as the midpoint between times $i$ and $i+1$ in discrete time. When $X$ is the computing quantity the predictive capacity quantifies the ability to predict the next step along the dashed blue line given knowledge of the solid blue and red lines. The capacity derived from the blue line is the active information storage whilst the capacity derived from the red line is the transfer entropy. The third subplot concerns the  the time reversed prediction associated with the time reversed computation. Here the reverse sequence of $X$ is being predicted about time $t$, however the prediction is in the same direction as the time evolution as the physical, time-reversed dynamics are being utilized. The fourth subplot serves to distinguish the reverse prediction from a \emph{retrodiction}: in this plot \emph{previous} values of $X$ are being retrodicted after the fact from the subsequent trajectories.
\par
 Importantly, as with the intrinsic computation, the time reversed predictive capacity naturally divides into two components
 \begin{align}
c^{\dagger,\{i,i+1\}}_x&=\ln\frac{p^\dagger(x_{i}|x_{n:i\text{+}1},y_{n:i\text{+}1})}{p(x_{i})}\nonumber\\
&=\ln\frac{p^\dagger(x_{i}|x_{n:i\text{+}1})}{p(x_{i})}+\ln\frac{p^\dagger(x_{i}|x_{n:i\text{+}1},y_{n:i\text{+}1})}{p^\dagger(x_{i}|x_{n:i\text{+}1})}.
\end{align}
The first characterises the uncertainty reduction due to the history of the variable which defines the computed quantity, $X$, and the second characterises uncertainty reduction due to the extraneous time series, $Y$.
\begin{figure*}[!htb]
\includegraphics[width=0.8\textwidth]{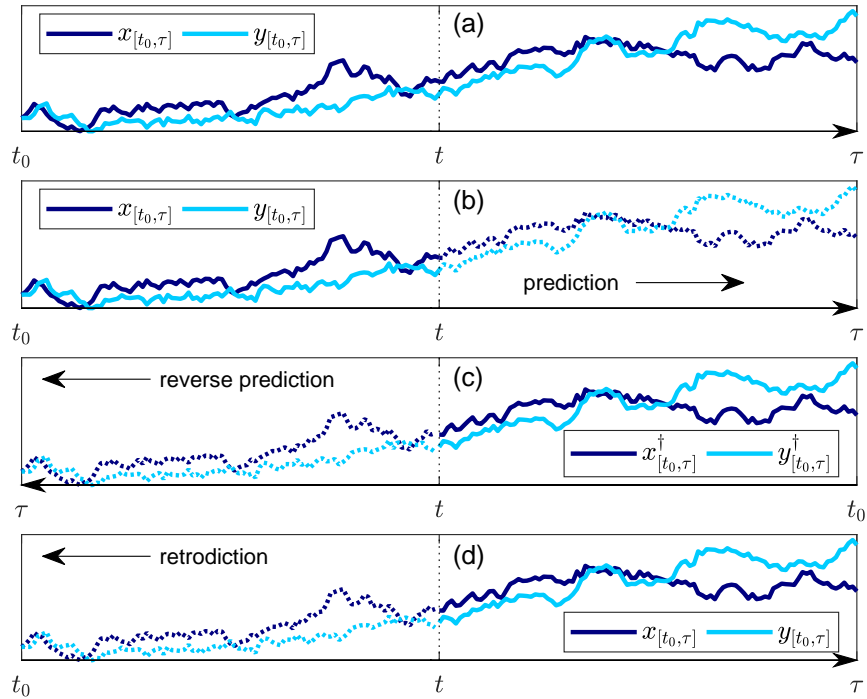}
\caption{\label{fig0} Illustration of the intrinsic and reverse computations/predictions. In the sub-figure (a) we plot two paths associated with random functions $X_{[t_0,\tau]}$ and $Y_{[t_0,\tau]}$ with time origin and horizon $t_0$ and $\tau$, with a specific intermediate point, $t_0< t<\tau$, being considered. At this time the forward (ensemble) computation in $X$ concerns the prediction of the dashed dark line on the right hand side in sub-figure (b) from the solid dark and light lines on the left hand side using the ensemble dynamics $p$. The reverse computation in $X$ concerns the prediction of the dashed dark line in sub-figure (c) from right to left using the time reversed dynamics $p^\dagger$. This is to be contrasted with retrodiction illustrated in sub-figure (d).}
\end{figure*}
The second quantity exists in the literature and is the local \emph{time reversed transfer entropy} \cite{spinney_transfer_2016}
\begin{align}
t^{\dagger,\{i,i\text{+}1\}}_{y\to x}&=\ln{\frac{p^\dagger(x_{i}|x_{n:i\text{+}1},y_{n:i\text{+}1})}{p^\dagger(x_{i}|x_{n:i\text{+}1})}},
\end{align} 
which becomes a pathwise quantity when integrated along a path \cite{spinney_transfer_2017}. The first quantity, however, has not previously been addressed and is a quantity we introduce which we call the local \emph{time reversed active information storage}
\begin{align}
a^{\dagger,\{i,i\text{+}1\}}_{ x}&=\ln{\frac{p^\dagger(x_{i}|x_{n:i\text{+}1})}{p(x_{i})}}.
\end{align} 
Together, these quantities characterise the distributed structure of time reversed computation and crucially do so on a \emph{local} scale for individual realisations of the process.
\section{A measure of computational irreversibility}
\label{comprev}
In quantifying the irreversibility of the computation we seek to contrast the computational signature of the intrinsic computation with the computational signature of the time reversed computation. Further, any such measure should characterise individual realisations of a process. We identify such a quantity as the difference between computational signatures, characterizing, on the local level, the difference in predictive capacity between the two predictions/computations. We introduce such a quantity, that we call the \emph{computational irreversibility}, as follows
\begin{align}
\mathcal{I}^{\{i,i+1\}}_{x}&=c^{\{i,i\text{+}1\}}_x-c^{\dagger,\{i,i\text{+}1\}}_x\nonumber\\
&=a^{\{i,i\text{+}1\}}_x-a^{\dagger,\{i,i\text{+}1\}}_x\nonumber\\
&\quad+t^{\{i,i\text{+}1\}}_{y\to x}-t^{\dagger,\{i,i\text{+}1\}}_{y\to x}.
\end{align}
where we identify components that contribute due to the predictive influence of $Y$ and $X$ as follows
\begin{align}
 \mathcal{I}^{\{i,i+1\}}_{x}&= \mathcal{I}^{A,\{i,i+1\}}_{x}+\mathcal{I}^{T,\{i,i+1\}}_{x\leftarrow y}\nonumber\\
 \mathcal{I}^{A,\{i,i+1\}}_{x}&=a^{\{i,i\text{+}1\}}_x-a^{\dagger,\{i,i\text{+}1\}}_x\nonumber\\
\mathcal{I}^{T,\{i,i+1\}}_{x\leftarrow y}&=t^{\{i,i\text{+}1\}}_{y\to x}-t^{\dagger,\{i,i\text{+}1\}}_{y\to x}.
\end{align}
As such we identify $\mathcal{I}^{A,\{i,i+1\}}_{x}$ as \emph{intrinsic} computational irreversibility associated with \emph{storage} of information in the transition from time $i$ to $i+1$ and $\mathcal{I}^{T,\{i,i+1\}}_{x\leftarrow y}$ as \emph{extrinsic} computational irreversibility associated with the \emph{transfer} of information over that transition.  Just as the active information storage quantifies the total predictability that can be extracted from the history of a time series, the intrinsic irreversibility quantifies the total irreversibility that can be identified using only that time series. Analogously, as the transfer entropy quantifies the additional predictability an extraneous time series provides, the extrinsic irreversibility quantifies the additional irreversibility that arises when $Y$ is known in addition to $X$.
\par
To proceed we need to associate such quantities with complete paths. As formulated, information dynamics considers its measures of storage and transfer to be  associated with individual time steps. Consequently we consider the difference in \emph{cumulatively} utilized information storage and information transfer along entire paths by considering the sum of those individual storage and transfer measures associated with each time step, here from a time origin $t=0$ to a time horizon $t=n$ obtaining, first, for the intrinsic irreversibility
\begin{align}
\Delta \mathcal{I}^{A,\{{0,n}\}}_{x}&=\sum_{i=0}^{n-1}  \mathcal{I}^{A,\{i,i+1\}}_{x}\nonumber\\
&=\sum_{i=0}^{n-1}a^{\{i,i\text{+}1\}}_x-a^{\dagger,\{i,i\text{+}1\}}_x\nonumber\\
&=\sum_{i=0}^{n-1}\ln\frac{p(x_{i+1}|x_{0:i})}{p(x_{i+1})}\frac{p(x_{i})}{p^\dagger(x_{i}|x_{n:i+1})}\nonumber\\
&=\ln{\frac{p(x_0)}{p(x_n)}}+\sum_{i=0}^{n-1}\ln{\frac{p(x_{i+1}|x_{0:i})}{p^\dagger(x_{i}|x_{n:i+1})}}\nonumber\\
&=\ln{\frac{p(x_{0:n})}{p^{\dagger}(x_{n:0})}}.
\end{align}
 Similarly the cumulative extrinsic irreversibility is defined as
\begin{align}
\Delta \mathcal{I}^{T,\{0,n\}}_{x\leftarrow y}&=\sum_{i=0}^{n-1} \mathcal{I}^{T,\{i,i+1\}}_{x\leftarrow y}\nonumber\\
&=\sum_{i=0}^{n-1}t^{\{i,i\text{+}1\}}_{y\to x}-t^{\dagger,\{i,i\text{+}1\}}_{y\to x}\nonumber\\
&=\sum_{i=0}^{n-1}\ln\frac{p(x_{i+1}|x_{0:i},y_{0:i})}{p(x_{i+1}|x_{0:i})}\frac{p^\dagger(x_{i}|x_{n:i+1})}{p^\dagger(x_{i}|x_{n:i+1},y_{n:i+1})}\nonumber\\
&=\ln{\frac{p^{\dagger}(x_{n\text{-}1:0}|x_n)}{p(x_{1:n}|x_0)}\frac{p(x_{1:n}|x_0,\{y_{0:n\text{-}1}\})}{p^{\dagger}(x_{n\text{-}1:0}|x_n,\{y_{n:1}\})}}
\end{align}
where we utilise notation for probabilities of complete paths as follows
\begin{align}
p(x_{1:n}|x_0)&=\prod_{i=1}^np(x_i|x_{0:i{-}1})\nonumber\\
p^\dagger(x_{n{-}1:0}|x_n)&=\prod_{i=0}^{n-1}p(x_i|x_{n:i{+}1})\nonumber\\
p(x_{0:n})&=p(x_0)p(x_{1:n}|x_0)\nonumber\\
p^\dagger(x_{n:0})&=p(x_n)p(x_{n-1:0}|x_n)\nonumber\\
p(x_{1:n}|x_0,\{y_{0:n\text{-}1}\})&=\prod_{i=0}^{n-1}p(x_{i+1}|x_{0:i},y_{0:i})\nonumber\\
p^\dagger(x_{n{-}1:0}|x_n,\{y_{n:1}\})&=\prod_{i=0}^{n-1}p(x_{i}|x_{n:i{+}1},y_{n:i{+}1})
\label{condnew}
\end{align}
with notation $\{\}$ emphasizing that $p(x_{1:n}|x_0,\{y_{0:n\text{-}1}\})\neq p(x_{1:n}|x_0,y_{0:n\text{-}1})$ as future values of $Y$ are causally blocked along evaluation of the path.
\par
Explicitly, $\Delta \mathcal{I}^{A,\{{0,n}\}}_{x}$ is the difference between the cumulative use of stored information utilized in a specific path for all computational steps from $t=0$ through to $t=n$ and the cumulative use of stored  information in all steps of the appropriate reverse computation sequenced from $n$ to $0$. Analogously $\Delta \mathcal{I}^{T,\{{0,n}\}}_{x\leftarrow y}$ is the difference between the cumulatively transferred information in the forward computations and the cumulatively transferred information in the appropriate reverse computations. Both of these are expressible as probabilities assigned to total path realisations, i.e. $x_{0:n}$ etc.
\par
Importantly, these accumulated quantities along such paths can be expressed as functionals consisting of ratios of the constructed path probabilities. This supports their existence when a limit in the discretization is taken in the approach to continuous time (thus permitting a rate). In the continuum regime we are concerned with such quantities formulated for continuous path functions written $x_\mathcal{A}\equiv\{x(t'):t'\in\mathcal{A}\}$.
We construct these quantities by considering an interval $[t_0,\tau]$ and constructing a discretisation running from $i=0$ to $i=n=(\tau-t_0)/\Delta t$ such that $x_i=x({t_0+i\Delta t})$ then taking $\Delta t\to 0$ such that $x_{0:n}\to x_{[t_0,\tau]}$, $x_{1:n}\to x_{(t_0,\tau]}$ and so on. In this limit, measures replace probabilities and computational signatures and irreversibilities become differential quantities defined instantaneously, i.e. $\mathcal{I}_x^{\{i,i+1\}}\to d\mathcal{I}_x(t)$. We consequently identify the accumulated quantities with logarithms of Radon-Nikodym derivatives between these path measures \cite{shargel_measuretheoretic_2010,spinney_transfer_2017}, such that for a process running from $t=t_0$ to $t=\tau$ we write
\begin{align}
\Delta \mathcal{I}^{A}_{x}&=\int_{t_0}^\tau d\mathcal{I}^{A}_x(t)\nonumber\\
&=\lim_{\Delta t\to 0}\ln{\frac{p(x_{0:n})}{p^{\dagger}(x_{n:0})}}\nonumber\\
&=\ln{\frac{dp[x_{[t_0,\tau]}]}{dp^{\dagger}[x_{[t_0,\tau]}]}}\nonumber\\
\Delta \mathcal{I}^{T}_{x\leftarrow y}&=\int_{t_0}^\tau d\mathcal{I}^{T}_{x\leftarrow y}(t)\nonumber\\
&=\lim_{\Delta t\to 0}\ln{\frac{p^{\dagger}(x_{n\text{-}1:0}|x_n)}{p(x_{1:n}|x_0)}\frac{p(x_{1:n}|x_0,\{y_{0:n\text{-}1}\})}{p^{\dagger}(x_{n\text{-}1:0}|x_n,\{y_{n:1}\})}}\nonumber\\
&=\ln{\frac{dp^{\dagger}[x_{(\tau,t_0]}|x_{\tau}]}{dp[x_{(t_0,\tau]}|x_{t_0}]}}\frac{dp[x_{(t_0,\tau]}|x_{t_0},\{y_{[t_0,\tau)}\}]}{dp^{\dagger}[x_{(\tau,t_0]}|x_{\tau},\{y_{[\tau,t_0)}\}]}.
\label{compprod}
\end{align}
For clarity, however, we utilise a deliberate abuse of notation rewriting measures as path probability `densities' such that
\begin{align}
\int dp[x_{[t_0,\tau]}] &\equiv\int dx_{[t_0,\tau]}\;p[x_{[t_0,\tau]}] \nonumber\\
&=\int dx_{[t_0,\tau]}\;p[x_{(t_0,\tau]}|x_{t_0}]p(x_{t_0}),
\end{align}
since this permits a more natural discussion and intuitive manipulation when dealing with familiar objects such as mutual informations etc. without altering any of the results.
\par
Before progressing we note that the intrinsic irreversibility associated with information storage permits a non-negative ensemble average viz.
\begin{align}
\langle \Delta \mathcal{I}^{A}_{x}\rangle&=\int dx_{[t_0,\tau]}\;p[x_{[t_0,\tau]}]\ln\frac{p[x_{[t_0,\tau]}]}{p^{\dagger}[x_{[\tau,t_0]}]}\geq 0,
\label{AS}
\end{align}
but no bound exists for $\langle \Delta \mathcal{I}_{x}\rangle$ or $\langle \Delta \mathcal{I}^{T}_{x\leftarrow y}\rangle$, properties that we shall return to.
\section{Computational irreversibility in physical systems}
The measures $\langle \Delta \mathcal{I}_{x}\rangle$, $\langle \Delta \mathcal{I}^A_{x}\rangle$ and $\langle \Delta \mathcal{I}^T_{x\leftarrow y}\rangle$ are very general objects which quantify irreversibility for any dynamics where the relevant probability measures can be defined. However, in order to understand their potential role in, and connection with, (stochastic) thermodynamics we must consider their behaviour in relevant physical systems.  This will necessarily involve a constraining of the nature of the systems considered. To understand the relevant constraints it is instructive to give a brief account of how irreversibility in physical systems can be identified thermodynamically as entropy production.
 \par
 \subsection{Entropy production in stochastic thermodynamics}
Here we give a brief account of how entropy production can be understood in terms of dynamic irreversibility in systems that obey such Markovian dynamics, thus giving a brief overview of the necessary features of stochastic thermodynamics. As with all thermodynamic descriptions, we consider a division of the universe into a system of interest with degrees of freedom that we are aware of and an environment that we are mostly unconcerned with and so remain ignorant of. If the system is small enough, this ignorance of the environment renders the behaviour of the system unpredictable, or stochastic due to our failure to track its precise evolution. Further, if we have performed such a division sensibly, the environmental degrees of freedom will be fast in comparison to the system allowing us to make an assumption that the induced probabilistic dynamics are Markov, but also the assumption that the environment is at equilibrium. In such situations a powerful principle arises known as \emph{local detailed balance} \cite{bauer_local_2015,bergmann_new_1955,colangeli_meaningful_2011,derrida_nonequilibrium_2007,lebowitz_stationary_1959,lebowitz_gallavotticohentype_1999,maes_time-reversal_2003,tasaki_two_2007} which relates the transition rates that arise from exposure to the environment to the exported entropy production realized in the environment. 
\par
 If we consider such a continuous time Markovian process in $X$ on a discrete state space we may describe its behaviour as a sequence of instantaneous transitions. On an interval $[t_0,\tau]$ there is a countable number of transitions, $N$, occurring at times $t_0<t_1<\ldots<t_{N}<\tau$ between states $X_{t}=x_{i}$, $t\in{[t_{i},t_{i+1})}$. The dynamics are consequently described by a set of transition rates $k^y_{x\to x'}$ concerning a transition from $x$ to $x'$ which we also allow to be parametrized by a smoothly varying switching protocol $Y$. In such a scenario any such transition is mediated by the equilibrium environment and thus the local detailed balance relation stipulates that it is accompanied by an entropy change in the environment, at time $t_{i}$, equal to
\begin{align}
\ln\frac{k^{y_{t_i}}_{x_{i-1}\to x_{i}}}{k^{y_{t_i}}_{x_{i}\to x_{i-1}}}=\Delta \mathcal{S}^x_{\rm env}(t_{i})
\end{align}
manifesting as a heat flow scaled by the environmental temperature when the environment is a heat bath. By definition, the probability of observing such a transition in a given time $\Delta t$ is then written $p(X_{t+\Delta t}=x'|X_t=x,Y_t=y)=k^y_{x\to x'}\Delta t+\mathcal{O}(\Delta t^2)$ such that we can associate the above with transition probabilities on a short time scale
\begin{align}
&\Delta \mathcal{S}^x_{\rm env}(t_{i})\nonumber\\
&=\ln\frac{k_{x_{i-1}\to x_{i}}}{k_{x_{i}\to x_{i-1}}}\nonumber\\
&=\lim_{\Delta t\to 0}\ln\frac{p(X_{t_{i}}=x_{i}|X_{t_{i}-\Delta t}=x_{i-1},Y_{t_{i}-\Delta t}=y_{t_i-\Delta t})}{p(X_{t_{i}}=x_{i-1}|X_{t_{i}-\Delta t}=x_{i},Y_{t_{i}-\Delta t}=y_{t_i})}.
\end{align}
If we have invariance in the time reversed dynamics $p^\dagger=p$ (i.e. $k^y_{x\to x'}=k^{y,\dagger}_{x\to x'}$) and $Y$ is suitably smooth then we may rewrite the above as 
\begin{align}
&\Delta \mathcal{S}^x_{\rm env}(t_{i})\nonumber\\
&=\lim_{\Delta t\to 0}\ln\frac{p(X_{t_{i}}=x_{i}|X_{t_{i}-\Delta t}=x_{i-1},Y_{t_{i}-\Delta t}=y_{t_i-\Delta t})}{p^\dagger(X_{t_{i}}=x_{i-1}|X_{t_{i}-\Delta t}=x_{i},Y_{t_{i}-\Delta t}=y_{t_i-\Delta t})}\nonumber\\
&=\lim_{\Delta t\to 0}\ln\frac{p(x_{t_i}|x_{t_i-\Delta t},y_{t_i-\Delta t})}{p^\dagger(x_{t_i-\Delta t}|x_{t_i},y_{t_i})},
\end{align}
i.e. in a form represented by the time reversed dynamics and path sequence.
\par
 In contrast, an absence of entropy production in the environment (due to $X$) is associated with the times between transitions, understood physically as an absence of exchange of energy with the environment. This too can be captured in terms of probabilities of the transition sequence by simply observing the probability of such an absence in the forward and time reversed dynamics to be exponentially distributed due to the Markovian dynamics and thus given by
\begin{align}
&\ln\frac{p(X_{(t_i,t_{i+1})}=x_i|X_{t_i}=x_i,\{Y_{[t_i,t_{i+1})}=y_{[t_i,t_{i+1})}\})}{p^\dagger(X_{[t_i,t_{i+1})}=x_i|X_{t_i}=x_i,\{Y_{[t_i,t_{i+1})}=y_{[t_{i+1},t_{i})}\})}\nonumber\\
&\qquad=\sum_{x'}\int_{t_i}^{t_{i+1}}(k^{y(t),\dagger}_{x_i\to x'}-k^{y(t)}_{x_i\to x'})dt,
\end{align}
which vanishes when $p=p^\dagger$ such that $k_{x\to x'}=k^\dagger_{x\to x'}$.
These contributions can be constructed in a piece-wise fashion, with waiting times interleaved between transitions, allowing us to consider the ratio of the conditional probability of the entire path $x_{[t_0,\tau]}$ and the conditional probability of the entire path $x_{[\tau,t_0]}$ under the time reversed dynamics allowing us to express the  total entropy exported by the system $X$ to the environment, along an entire path $x_{[t_0,\tau]}$ as \cite{seifert_entropy_2005,esposito_three_2010}
\begin{align}
\Delta \mathcal{S}^x_{\rm env}[x_{[t_0,\tau]}]&=\int_{t_0}^\tau d\mathcal{S}_{\rm env}^x(t)\nonumber\\
&=\ln\frac{p[x_{(t_0,\tau]}|x_{t_0},\{y_{[t_0,\tau)}\}]}{p^{\dagger}[x_{(\tau,t_0]}|x_{\tau},\{y_{[\tau,t_0)}\}]}
\label{eq25}
\end{align}
in this case with $d\mathcal{S}_{\rm env}^x(t)=\delta(t-t_i)\Delta\mathcal{S}_{\rm env}(t_i)$. 
Whilst we have described the above for a jump process, the above relationship is very robust, holding for all Markov systems that possess a local detailed balance relationship. This can be understood by recognizing that continuous dynamics, realized, for instance, as stochastic differential equations, may be achieved through the appropriate limit of such a jump process (i.e. the Fokker Planck equation may be derived from the master equation).
\par
Having established the form of the exported entropy production,  one of the key insights of stochastic thermodynamics is to observe that when augmenting such a quantity with a change in \emph{pointwise} (or \emph{local}) Shannon entropy of the system, the sum then represents the total entropy production, in model, of the universe \cite{seifert_entropy_2005} (consisting of the system and the environment) and obeys the requisite statistical requirements to satisfy a second law. Such a change in Shannon entropy,
for any class of system on the interval $[t_0,\tau]$ reads
\begin{align}
\Delta\mathcal{S}^x_{\rm sys}[x_{[t_0,\tau]}]&=\ln\frac{p(x_{t_0})}{p(x_{\tau})}.
\end{align}
If we consider such a change between the start and end of the process on $x_{[t,\tau]}$ this provides an initial condition for the path probabilities appearing in  Eq.~(\ref{eq25}) such that
\begin{align}
\Delta \mathcal{S}^x_{\rm tot}[x_{[t_0,\tau]},y_{[t_0,\tau]}]&=\Delta \mathcal{S}^x_{\rm sys}[x_{[t_0,\tau]},y_{[t_0,\tau]}]\nonumber\\
&\qquad+\Delta \mathcal{S}^x_{\rm med}[x_{[t_0,\tau]},y_{[t_0,\tau]}]\nonumber\\
&=\ln\frac{p[x_{[t_0,\tau]}|x_{t_0},\{y_{[t_0,\tau)}\}]p(x_{t_0})}{p^{\dagger}[x_{[\tau,t_0]}|x_{\tau},\{y_{[\tau,t_0)}\}]p(x_{\tau})}\nonumber\\
&=\ln\frac{p[x_{[t_0,\tau]}|\{y_{[t_0,\tau]}\}]}{p^{\dagger}[x_{[\tau,t_0)}|\{y_{[\tau,t_0)}\}]}.
\label{entprod}
\end{align}
It is this quantity that the second law governs, and it does so on a statistical level. As  such, assuming no feedback between the system $X$ and switching protocol $Y$ (i.e. $Y$ cannot depend on $X$), the following holds
\begin{align}
\langle\Delta\mathcal{S}^x_{\rm tot}\rangle=\langle\Delta\mathcal{S}^x_{\rm sys}\rangle+\langle\Delta\mathcal{S}^x_{\rm env}\rangle\geq 0
\end{align}
representing the second law.
\par
Finally, however, it is important to note two key assumptions. The first is the assumption of no feedback. If this is not met then the second law as written above need not hold. To recover the second law in such cases either the definition of the system must be expanded or the result generalised. The second is the assumption of smoothness in $Y$ in the preceding identification of the entropy production and its representation in terms of dynamical probabilities. If such a criterion is not met the notion of a heat physically dissipated by $X$ can become fundamentally ambiguous since the value $y$ that enters the local detailed balance condition is not well defined. This can occur notably if i) both $X$ and $Y$ are continuous but nowhere differentiable (as in the case of Langevin equations) and driven by correlated noise or ii) both $X$ and $Y$ are discrete variables which can simultaneously transition. In the latter, such a situation can be avoided by insisting that joint transitions are disallowed such that the dynamics of \emph{both} $X$ and $Y$, governed by the transition rates $k_{x\to x'}^{y\to y'}$ ($\{x,y\}\neq \{x',y'\}$) may be written
\begin{align}
k_{x\to x'}^{y\to y'}=(1-\delta_{x,x'})\delta_{y,y'}k_{x\to x'}^y+(1-\delta_{y,y'})\delta_{x,x'}k_{x}^{y\to y'}
\end{align}
such that $\sum_{y'}k_{x\to x'}^{y\to y'}=k_{x\to x'}^y$ and $\sum_{x'}k_{x\to x'}^{y\to y'}=k_{x}^{y\to y'}$. 
More broadly such a requirement specifies conditional independence in the subsequent values of $X$ and $Y$ given the preceding history over small time scales, arising here since the transition times of $X$ and $Y$ are not correlated as they do not transition together, almost surely. This conditional independence is expressed by the following
\begin{align}
&\lim_{dt\to 0}p(x_{t+dt},y_{t+dt}|x_t,y_t)\nonumber\\ 
&\qquad= \lim_{dt\to 0}p(x_{t+dt}|x_t,y_t)p(y_{t+dt}|x_t,y_t)
\label{bipbip}
\end{align}
which we can observe to hold for the discrete space case by considering a transition $x\to x'\neq x$, $y=y'$, without loss of generality, that  we have
\begin{align}
&p(X_{t+dt}=x'|X_t=x,Y_t=y)p(Y_{t+dt}=y|X_t=x,Y_t=y)\nonumber\\
&\qquad=k_{x\to x'}^ydt(1-\sum_{y'}k_{x}^{y\to y'}dt)+\mathcal{O}(dt^2)\nonumber\\
&\qquad=k_{x\to x'}^y dt+\mathcal{O}(dt^2)\nonumber\\
&\qquad=p(X_{t+dt}=x',Y_{t+dt}=y|X_t=x,Y_t=y)+\mathcal{O}(dt^2).
\end{align}
In the former case with continuous paths such a conditional independence requirement manifests as statistical independence of any noise terms between $X$ and $Y$. When such a requirement is met then one may write
\begin{align}
\ln\frac{p(x_{t+dt}|x_t,y_t)}{p^\dagger(x_{t}|x_{t+dt},y_{t+dt})}=\ln\frac{p(x_{t+dt}|x_t,y_t)}{p^\dagger(x_{t}|x_{t+dt},y_{t})}+\mathcal{O}(dt^2),
\end{align}
identifying a term representing the local detailed balance condition on the right hand side and recovering Eq.~(\ref{eq25}).
\par
This property of conditional independence, manifest as independence in noise terms in continuous dynamics and illegality of joint transitions in discrete dynamics, is called a \emph{bipartite} property.
\subsection{Relating computational irreversibility and physical entropy production }
Here we make an explicit connection between the measures of computational irreversibility established in section \ref{comprev} and the thermodynamic quantities described above. To do so we consider, first, a system where the physical assumptions listed above are true, namely that: i) an equilibrium environment leads to stochastic dynamics in a system, $X$, which are Markov thus leading to a local detailed balance relation and ii) that the joint dynamics of the system $X$ and some extraneous parameter $Y$, which may represent an external protocol or some other coupled subsystem, are bipartite such that we observe no joint transitions or consider independence of noise in the continuous limit. 
\par
In this case we straightforwardly compare the contents of Eq.~(\ref{entprod}) and Eq.~(\ref{compprod}) and and immediately recognize a central observation
\begin{align}
\Delta\mathcal{I}_x&=\Delta\mathcal{S}^x_{\rm tot}.
\end{align}
Explicitly, the computational irreversibility reduces to the total thermodynamic entropy production in the case of Markovian bipartite dynamics  revealing that the central measure of intrinsic computation, the predictive capacity, can be seen to be an implicitly central quantity in stochastic thermodynamics. In other words, entropy production can be equivalently described by physical irreversibility and equivalently by computational irreversibility as measured by information dynamics.
\par
Since thermodynamic and computational irreversibility are equivalent in such a setting and the computational irreversibility is designed around a division in storage and transfer of information, it is instructive to translate such a division, realized in the intrinsic and extrinsic computation irreversibilities in Eq.~(\ref{compprod}), into divisions of physical entropy production. As such we therefore explicitly write
\begin{align}
d\mathcal{S}^x_{\rm tot}&=d\mathcal{S}^x_{\rm storage}+d\mathcal{S}^{x\leftarrow y}_{\rm transfer}
\label{div3}
\end{align}
where we identify $d\mathcal{S}^x_{\rm storage}=d\mathcal{I}^{A}_{x}$ as the \emph{physical entropy production} associated with the active storage of information and $d\mathcal{S}^{x\leftarrow y}_{\rm transfer}=d\mathcal{I}^{T}_{x\leftarrow y}$ as the physical entropy production associated with the transfer of information based on the identification $d\mathcal{S}^x_{\rm tot}=d\mathcal{I}_x=d\mathcal{I}^{A}_{x}+d\mathcal{I}^{T}_{x\leftarrow y}$.  The last quantity, $d\mathcal{S}^{x\leftarrow y}_{\rm transfer}$ was used in the construction of a measure of information theoretic time asymmetry, an `information theoretic arrow of time' \cite{spinney_transfer_2016}.
We emphasize here, however, that we are making a complete division of the entropy production associated with the computational primitives, deriving from general measures of irreversibility.
\par
Such a division is to be contrasted with other important divisions of the total entropy production of the universe, such as the usual division into the entropy change attributed to the system and the environment \cite{seifert_entropy_2005} (or in the language of Prigogine, the system into the internal and external productions \cite{i._prigogine_thermodynamic_1967})
\begin{align}
d\mathcal{S}^x_{\rm tot}&=d\mathcal{S}^x_{\rm sys}+d\mathcal{S}^x_{\rm env}
\label{div1}
\end{align}
or the more recent division of the total entropy production into adiabatic and non-adiabatic contributions \cite{chetrite_fluctuation_2008,esposito_three_2010,esposito_three_2010-1,spinney_nonequilibrium_2012,vandenbroeck_three_2010}
\begin{align}
d\mathcal{S}^x_{\rm tot}&=d\mathcal{S}^x_{\rm na}+d\mathcal{S}^x_{\rm a}.
\label{div2}
\end{align} 
All of the above divisions of the total entropy production are based, in some sense, on how they manifest. The division in Eq.~(\ref{div1}) characterises the \emph{location} associated with the entropy change: $d\mathcal{S}^x_{\rm sys}$ concerns the entropy change associated with the system whilst $d\mathcal{S}^x_{\rm env}$ concerns the entropy change of the environment. The division in Eq.~(\ref{div2}) characterises the ensemble level \emph{mechanisms} that lead to dissipation: $d\mathcal{S}^x_{\rm na}$ concerns the entropy production associated with driving \& relaxation and $d\mathcal{S}^x_{\rm a}$ concerns the entropy production associated with steady non-equilibrium constraints. Our division in Eq.~(\ref{div3}) does so based on the their manifestation of computational behaviour (storage and transfer), defined in terms of \emph{predictive capacity} based on the terms laid out in section \ref{infodyn}.
\par
Finally, we mention that since the total computational irreversibility can be associated with physical entropy production, by assuming a heat bath at constant temperature $\beta^{-1}$ (we use $k_B=1$ throughout), we can write
\begin{align}
d\mathcal{S}^x_{\rm storage}&=d\mathcal{S}_{\rm sys}+d\beta Q^x_{\rm storage}\nonumber\\
d\mathcal{S}^{x\leftarrow y}_{\rm transfer}&=d\beta Q^{x\leftarrow y}_{\rm transfer}
\end{align}
where
\begin{align}
\Delta Q^x=\Delta Q^x_{\rm storage}+\Delta Q^{x\leftarrow y}_{\rm transfer}.
\label{divQ}
\end{align}
reminiscent of the identification of the non-adiabatic entropy production with the system entropy and excess heat \cite{esposito_three_2010,hatano_steadystate_2001}, and the adiabatic entropy production with the house keeping heat \cite{esposito_three_2010,speck_integral_2005,oono_steady_1998}. That is we can identify contributions to physical heat flows associated with storage and transfer of information.
\par
Next, we turn our attention to the behaviour of these contributions to the total entropy production. Firstly we consider the situation outlined above, that of Markovian bipartite dynamics before considering how they behave when the bipartite assumption is dropped.
\section{Bipartite systems}
In this section we explore the behaviour of the entropy productions associated with the storage and transfer of information in systems that are described through the dynamics of two variables $X$ and $Y$ and where the dynamics are of these variables are both bipartite and Markov. To reiterate, these are systems where, if jumps in the variables are permitted, they cannot occur simultaneously in both $X$ and $Y$. Continuous bipartite dynamics can be considered by considering the limiting behaviour of such systems resulting in a conditional independence condition.
\par
It is important to understand that such dynamics may describe both a system comprising two components of a total physical system in contact with a heat reservoir, or a single variable $X$ controlled by switching protocol $Y$ as the dynamic assumptions in both cases are identical. As such in this section we first consider the contributions associated with system $X$ in the context of of a protocol $Y$, before describing the joint contributions of the composite system.
\subsection{Single variable and switching protocol}
Here, we consider the simplest systems available for consideration. These systems are the typical, canonical, systems found in the original literature of stochastic thermodynamics consisting of a single particle or object of interest denoted the `system' (which may be a composite system itself), represented here by $X$, and a switching protocol that determines the nature of the dynamics through the change of a potential/nature of a heat bath etc., represented here by $Y$. 
In addition to the Markov and bipartite dynamics, and in contrast to systems we consider subsequently, we do not permit any feedback between the switching protocol and the particle. Consequently it follows that we may write the probability of observing the joint path $\{x_{[t_0,\tau]},y_{[t_0,\tau]}\}$
\begin{align}
&p[x_{[t_0,\tau]},y_{[t_0,\tau]}]\nonumber\\
&=p(x_{t_0},y_{t_0})p[x_{(t_0,\tau]}|x_{t_0},\{y_{[t_0,\tau)}\}]p[y_{(t_0,\tau]}|y_{t_0},\{x_{[t_0,\tau)}\}]\nonumber\\
&=p[x_{(t_0,\tau]}|x_{t_0},\{y_{(t_0,\tau]}\}]p[y_{(t_0,\tau]}|y_{t_0}]p(x_{t_0}|y_{t_0})p(y_{t_0})\nonumber\\
&=p[x_{[t_0,\tau]}|y_{[t_0,\tau]}]p[y_{[t_0,\tau]}],
\end{align}
noting that here the evolution of $Y$ is independent of $X$, i.e. $p[y_{(t_0,\tau]}|y_{t_0},\{x_{[t_0,\tau)}\}]=p[y_{(t_0,\tau]}|y_{t_0}]$, and thus that $p[x_{(t_0,\tau]}|x_{t_0},\{y_{(t_0,\tau]}\}]=p[x_{(t_0,\tau]}|x_{t_0},y_{(t_0,\tau]}]$. 
\par
The division in Eq.~(\ref{divQ}) identifies heat flows based on information theoretic constructs, on the level of individual behaviour,  provided by the physical dynamics, but also by the statistics of the ensemble which serves to change the character of the predictability of the process. Explicitly this means changing the properties of the ensemble amounts to changing the amount of information that can be deemed to be in active storage or being transferred and thus the irreversibility associated with it. 
\par
As such it is illustrative to consider different situations where one observes different storage and transfer behaviour and thus different contributions to the total entropy production. To do so under the system/protocol paradigm without feedback we consider three cases, each with increasing constraint. 
\subsubsection{Irreversible protocols}
First we consider the case where the protocol is chosen randomly according to $p[y_{[t_0,\tau]}]$, and not necessarily reversibly, i.e. $p[y_{[t_0,\tau]}]\neq p^\dagger[y_{[\tau,t_0]}]$. In this case we have, in addition to $\langle \Delta\mathcal{S}^x_{\rm storage}\rangle\geq 0$ (which  emerges from Eq.~(\ref{AS})) an integral fluctuation theorem regarding the total entropy production in $X$, plus a change in correlation between system and protocol over the process (captured by a change in a local mutual information, $\Delta i_M$, where $\langle i_M\rangle =I_M$ and $i_M=\ln[p(x,y)/(p(x)p(y))]$). Specifically we may write
\begin{align}
&\langle\exp[-\Delta\mathcal{S}^x_{\rm storage}-\Delta\mathcal{S}^{x\leftarrow y}_{\rm transfer}+\Delta i_M]\rangle=\nonumber\\
&\int dx_{[t_0,\tau]}\int dy_{[t_0,\tau]}\nonumber\\
&\qquad\times p[y_{[t_0,\tau]}]p[x_{(t_0,\tau]}|x_{t_0},\{y_{[t_0,\tau)}\}]p(x_{t_0}|y_{t_0})\nonumber\\
&\qquad\times\frac{p^\dagger[x_{(\tau,t_0]}|x_\tau,\{y_{[\tau,t_0)}\}]p(x_\tau)p(x_{t_0})p(x_\tau|y_\tau)}{p[x_{(t_0,\tau]}|x_{t_0},\{y_{[t_0,\tau)}\}]p(x_{t_0})p(x_{t_0}|y_{t_0})p(x_\tau)}\nonumber\\
&=\int dy_{[t_0,\tau]}\;p[y_{[t_0,\tau]}]\nonumber\\
&\qquad\times\int dx_{[t_0,\tau]} p^\dagger[x_{(\tau,t_0]}|x_\tau,\{y_{[\tau,t_0)}\}]p(x_\tau|y_\tau)\nonumber\\
&=1
\end{align}
such that we have $\langle \Delta\mathcal{S}^x_{\rm storage}+\Delta\mathcal{S}^{x\leftarrow y}_{\rm transfer}\rangle-\Delta I_M\geq 0$ following from Jensen's inequality. Consequently we see that we have the bound
\begin{align}
\langle \Delta\mathcal{S}^{x\leftarrow y}_{\rm transfer}\rangle\geq \Delta I_M-\langle\Delta\mathcal{S}^x_{\rm storage}\rangle
\end{align}
or rather
\begin{align}
\Delta S_x -\Delta I_M +\beta\langle\Delta Q^x_{\rm storage}\rangle+\beta\langle\Delta Q^{x\leftarrow y}_{\rm transfer}\rangle\geq 0
\label{bound1}
\end{align}
where $\Delta S_x=\langle\Delta\mathcal{S}^x_{\rm sys}\rangle$. We note that the final mutual information, is not strictly necessary for the relation Eq.~(\ref{bound1}) to hold, meaning that for an initially uncorrelated system and protocol, we find
\begin{align}
\langle \Delta\mathcal{S}^{x\leftarrow y}_{\rm transfer}\rangle\geq-\langle\Delta\mathcal{S}^x_{\rm storage}\rangle.
\end{align}
Since $\langle\Delta\mathcal{S}^x_{\rm storage}\rangle\geq 0$, $\langle \Delta\mathcal{S}^{x\leftarrow y}_{\rm transfer}\rangle$ \emph{can} be negative, but not enough so that the total entropy production becomes negative. This forms a central behaviour of the two contributions to entropy production: the intrinsic entropy production due to storage is non-negative, yet the extrinsic entropy production due to transfer may take negative values, potentially offsetting contributions from its surroundings. The above result points out that even when the total entropy production (their sum) is non-negative the entropy due to transfer may still be negative.
\subsubsection{Reversible protocols}
To see why $\langle \Delta\mathcal{S}^{x\leftarrow y}_{\rm transfer}\rangle$ can take negative values, it is instructive to consider progressively more restrictive processes. As such we consider a process where there is no irreversibility in the protocol: this can be considered to mean that the evolution of the protocol is purely diffusive, in both the forward and reverse processes, or rather that the protocol in the reverse process is guaranteed to be the time reversed protocol of the forward process. Alternatively, one may simply posit that protocols are chosen from a distribution which is symmetric with respect to time reversal. All, however, yield the same condition, $p[y_{[t_0,\tau]}]=p^\dagger[y_{[\tau,t_0]}]$. When such a condition holds, in addition to the positivity of the storage entropy production $\langle \Delta\mathcal{S}^x_{\rm storage}\rangle\geq 0$, $\langle \Delta\mathcal{S}^x_{\rm sys}+\beta \Delta Q^x_{\rm storage}\rangle\geq 0$ following from Eq.~(\ref{AS}), we may write, 
\begin{align}
&\langle\exp[\Delta i_M-\Delta\mathcal{S}^{x\leftarrow y}_{\rm transfer}]\rangle\nonumber\\
&=\int dx_{[t_0,\tau]}\int dy_{[t_0,\tau]}\nonumber\\
&\qquad\times p(x_{t_0}|y_{t_0})p[x_{(t_0,\tau]}|x_{t_0},\{y_{[t_0,\tau)}\}]p[y_{[t_0,\tau]}]\nonumber\\
&\qquad\times\frac{p(x_{t_0})}{p(x_{\tau})}\frac{p(x_\tau|y_\tau)}{p(x_{t_0}|y_{t_0})}
\frac{p[x_{(t_0,\tau]}|x_{t_0}]}{p[x_{(t_0,\tau]}|x_{t_0},\{y_{[t_0,\tau)}\}]}
\nonumber\\
&\qquad\times \frac{p^\dagger[x_{(\tau,t_0]}|x_\tau,\{y_{[\tau,t_0)}\}]}{p^\dagger[x_{(\tau,t_0]}|x_\tau]}\nonumber\\
&=\int dx_{[t_0,\tau]}\int dy_{[t_0,\tau]}\;p(x_{\tau}|y_{\tau})p^\dagger[x_{(\tau,t_0]}|x_\tau,\{y_{(\tau,t_0]}\}]\nonumber\\
&\qquad\times p[y_{[t_0,\tau]}]\frac{p[x_{[t_0,\tau]}]}{p^{\dagger}[x_{[\tau,t_0]]}]}\nonumber\\
&=\int dx_{[t_0,\tau]}\int dy_{[t_0,\tau]}\;p(x_{\tau}|y_{\tau})\nonumber\\
&\qquad\times p^\dagger[x_{(t_0,\tau]}|x_\tau,\{y_{[t_0,\tau)}\}]p^\dagger[y_{[\tau,t_0]}]\frac{p[x_{[t_0,\tau]}]}{p^\dagger[x_{[\tau,t_0]}]}\nonumber\\
&=\int dx_{[t_0,\tau]}\;p^\dagger[x_{[\tau,t_0]}]\frac{p[x_{[t_0,\tau]}]}{p^\dagger[x_{[\tau,t_0]}]}\nonumber\\
&=1
\end{align}
such that, again, by Jensen's inequality, 
\begin{align}
-\Delta I_M+\langle\Delta\mathcal{S}^{x\leftarrow y}_{\rm transfer}\rangle&=-\Delta I_M+\langle \beta \Delta Q^{x\leftarrow y}_{\rm transfer}\rangle\geq 0.
\end{align}
I.e. in this picture the single relation of Eq.~(\ref{bound1}) decomposes into two distinct relations which we may write as two Clausius-like statements for these processes, namely
\begin{align}
\Delta S_x &\geq -\beta \int_{t_0}^t\langle dQ^x_{\rm storage}\rangle\nonumber\\
-\Delta I_M &\geq -\beta \int_{t_0}^t\langle dQ^{x\leftarrow y}_{\rm transfer}\rangle.
\label{bound2}
\end{align}
Differences in uncertainty and correlation, bound heat flows associated with storage and transfer of information respectively. I.e. both reflect heat flows that are associated with behaviour attributable to $X$ and its interaction with extrinsic variables. Crucially, however, we must understand that it is the absence of irreversibility in the extrinsic process has lead to the decomposition into two separate bounds facilitated by the stronger bound on $\langle \Delta\mathcal{S}^{x\leftarrow y}_{\rm transfer}\rangle$. In other words, there is no irreversibility extrinsic to $X$ which the entropy due to information transfer need offset and thus the additional bound emerges.
\subsubsection{Deterministic protocols}
Finally, for completeness, we can consider the more traditional conception of a process driven by a switching protocol, namely that there is a single protocol $y_{[t_0,\tau]}=\vec{y}$, which can be reliably selected, repeatedly, such that we have $p[y_{[t_0,\tau]}]=\delta(y_{[t_0,\tau]}-\vec{y})$. This in turn leads to $p[x_{(t_0,\tau]}|x_{t_0}]\to p[x_{(t_0,\tau]}|x_{t_0},\{\vec{y}\}]$ and thus $\Delta \mathcal{S}^{x\leftarrow y}_{\rm transfer}=\Delta \mathcal{S}^{x\leftarrow \emptyset}_{\rm transfer}=0$ for all realisations. This simply serves to illustrate the fact that information is only deemed to be in transfer when there is a non-singular ensemble associated with the extrinsic behaviour, such that the division of entropy into stored and transferred components is trivial here.
\par
In summary, we see that when there is no variation in the extrinsic process, no transfer of information occurs and thus no entropy production can be associated with transfer of information since knowledge of the protocol adds no predictive capacity to the underlying computations. When this condition is relaxed, but we assert that the extrinsic process is perfectly reversible, the entropy due to transfer of information behaves like an additional physical heat flow, leading to two distinct second laws, as it is rigorously bounded by a change in uncertainty in the process. Finally we see that this second bound is lost when the dynamics of the protocol can be intrinsically irreversible. 
\par
The mechanism by which this happens is important. When the protocol is irreversible, on average the realized $y_{[t_0,t]}$ is more probable than the corresponding reverse protocol, $y_{[t,t_0]}$. By definition, the forward protocol is a less surprising event in the ensemble that characterises the marginalized dynamics $p[x_{(t_0,t]}|x_{t_0}]$ than the reverse protocol. This indicates that the reverse protocol has more predictive power in the reverse process because it is \emph{rarer} than the forward protocol. This in turn leads to a reduced entropy production arising from the transfer of information: one gains more predictive power using $Y$ in the reverse process than in the forward process \emph{precisely because it is rarer}. This property, where the entropy production due to transfer of information is (negatively) sensitive to the irreversibility of the extrinsic processes is a feature that underlies much of its behaviour in more complicated processes.
\par
Finally we note in this section we have introduced boundary mutual information terms, required for the relevant integral fluctuation theorems to hold, in a seemingly \emph{ad-hoc} and unprincipled way. Indeed similar terms appear in the literature \cite{ito_information_2013}. In the next section we offer a rationalization for the inclusion of such terms based on a principle of conservation of predictive capacity. 
\subsection{Bipartite systems with feedback}
Next we extend the systems considered above to systems where $X$ and $Y$, whilst remaining bipartite, may evolve dependently on each other such that effect feedback. This allows one to consider models of protocols ($Y$) which can respond to the system ($X$) or to consider a joint, composite, system, $\{X,Y\}$, identifying contributions to the entropy production from each.
\par
Especially relevant to the latter case is the fact that we may observe the evolution on different levels of description, each of which on has distinct characterisations both in terms of their thermodynamics and their intrinsic computation. So far we have only discussed the intrinsic computation and thermodynamics of individual subsystems, but to understand the role of entropy exchange in terms of intrinsic computation in such systems we must also understand those behaviours on the level of the composite system.
\par
To do so we must discuss the intrinsic computation, computational irreversibility and thus entropy production associated with the joint system. The joint system is one with only an intrinsic component and thus has a predictive capacity equal to its information storage, ie.
\begin{align}
c_{xy}=a_{xy}.
\end{align} 
We now seek to relate this predictive capacity to the information processing that is occurring on the level of the individual subsystems characterized through $c_x$ and $c_y$. To do we posit a principle that any coherent characterisation of the composite system should assign predictive capacity to sub-components in such a way so as to conserve the total of the joint system. To this end we write
\begin{align}
c_{xy}&= c_{x}+c_{y}+c^{xy}_{\rm{int}}
\label{cint}
\end{align}
where 
\begin{align}
c_{x}&=a_{x}+t_{y\to x}\nonumber\\
c_{y}&=a_{y}+t_{x\to y}
\end{align}
are the predictive capacities of the individual subsystems. We emphasize the appearance of $c^{xy}_{\rm{int}}$, a new quantity, which we call the the \emph{interaction prediction}. This quantifies the predictive power that emerges only when jointly predicting both $X$ and $Y$, i.e. the portion of the computational signature of the joint process which cannot be uniquely attached to any individual sub-system and arises from this necessity of conserving predictive capacity.  Importantly, unlike $c_{xy}$, $c_x$, $c_y$ or their constituent quantities, $c_{\rm int}^{xy}$ has, in general, no bound on its sign in expectation \emph{except} for bipartite systems where it must be negative. This occurs for bipartite systems since in this case $c_{\rm int}^{xy}$ is given by the negative pointwise mutual information
\begin{align}
c^{xy}_{\rm{int}}&=i_M=-\ln\frac{p(x,y)}{p(x)p(y)}.
\end{align}
Further, to relate such a decomposition to the computational irreversibilities and entropy production contributions we must attribute an irreversibility term to this interaction prediction. Using the definition of time reversed computations laid out in section \ref{revcomp} we may identify an appropriate \emph{time reversed interaction prediction}
\begin{align}
c_{xy}^{\rm{int},\dagger}&=c_{xy}^\dagger-c^\dagger_{x}-c^\dagger_{y}
\end{align}
which analogously leads to an \emph{interaction irreversibility}
\begin{align}
d\mathcal{I}_{xy}^{\rm int}&=c_{xy}^{\rm{int}}-c_{xy}^{\rm{int},\dagger}.
\end{align}
Since in these systems we identify $d\mathcal{I}_{xy}=d\mathcal{S}^{xy}_{\rm tot}$ and $d\mathcal{I}_{x}=d\mathcal{S}^{x}_{\rm tot}$, we identify the interaction irreversibility with an \emph{interaction entropy production}, $d\mathcal{I}_{xy}^{\rm int}=d\mathcal{S}_{\rm int}^{xy}$. Importantly, for these bipartite structures this simply reduces to the negative change in pointwise mutual information
\begin{align}
\Delta\mathcal{S}^{xy}_{\rm int}&=-\Delta i_M\nonumber\\
\langle\Delta\mathcal{S}^{xy}_{\rm int}\rangle&=-\Delta I_M.
\end{align}
This then provides a coherent framework for the inclusion of the mutual information in Eqs.~(\ref{bound1}) and (\ref{bound2}): it is accounting for computational irreversibility which cannot be associated with either $X$ or $Y$. Moreover, we shall see that such a principle forms the basis of a general approach to systems where fewer assumptions about the dynamics hold.
\par
Proceeding, as mentioned, a particular benefit of the bipartite assumption is that it explicitly allows us to establish heat flows associated with each component of the system, since the conditional independence of the transition probabilities leads to a joint local detailed balance condition decomposing into separate contributions from  both $X$ and $Y$ \cite{horowitz_thermodynamics_2014} such that we can write
\begin{align}
\Delta \mathcal{S}_{\rm tot}^{xy}&=\Delta\mathcal{S}^{xy}_{\rm int}+\Delta S_{\rm tot}^{x}+\Delta S_{\rm tot}^{y}\nonumber\\
&=-\Delta I_M+\Delta S_{\rm storage}^{x}+\Delta S_{\rm transfer}^{x\leftarrow y}\nonumber\\
&\qquad+\Delta S_{\rm storage}^{y}+\Delta S_{\rm transfer}^{y\leftarrow x}\nonumber\\
&=-\Delta I_M+\Delta S_{\rm sys}^{x}+\beta \Delta Q^x_{\rm storage}+\beta \Delta Q^{x\leftarrow y}_{\rm transfer}\nonumber\\
&\qquad+\Delta S_{\rm sys}^{y}+\beta \Delta Q^y_{\rm storage}+\beta \Delta Q^{y\leftarrow x}_{\rm transfer}
\end{align}
with the division of heat flows obeying
\begin{align}
\Delta Q^{xy}&=\Delta Q^{x}+\Delta Q^{y}\nonumber\\
&=\Delta Q^x_{\rm storage}+ \Delta Q^{x\leftarrow y}_{\rm transfer}+ \Delta Q^y_{\rm storage}+\Delta Q^{y\leftarrow x}_{\rm transfer}.
\label{bipheat}
\end{align}
Importantly, however, no individual contribution, e.g. $\Delta S_{\rm tot}^{x}=\Delta S_{\rm storage}^{x}+\Delta S_{\rm transfer}^{x\leftarrow y}$, nor, even sums of contributions, $\Delta S_{\rm tot}^{x}+\Delta S_{\rm tot}^{y}$, has a bound on its sign in expectation; in the context of multi-component systems the irreversibility of one component can be negative if it is commensurately offset by positive irreversibility in the remaining components. The second law applies to the total system (and environment), not necessarily on each component in isolation. Importantly, however, the remaining components which offset such a reduction \emph{must include the interaction entropy production}.
\par
The precise nature of how this phenomenon manifests has attracted much attention in terms of information flows, having its origins in measurement and feedback systems, identifying minimal correlations that can be exploited to reduce the entropy of a subsystem. In this work we take a different approach and identify the \emph{physical entropy} that can be offset in different subsystems, arguing that it is the entropy production due to information transfer that can cause negative entropy productions and indeed that it is the entropy production due to information transfer associated with other subsystems that have the ability to offset this reduction. The bounds that then arise from the these interactions then work in interplay with the interaction entropy production.
\par
Firstly we establish that the total entropy of the entire system is, necessarily, positive in expectation in accordance with the second law, formulated here as a contribution entirely associated with information storage and thus is rigorously non-negative by Eq.~(\ref{AS}), i.e.
\begin{align}
\langle \Delta \mathcal{S}_{\rm tot}^{xy}\rangle &= \int dx_{[t_0,\tau]}\int dy_{[t_0,\tau]}\; p[x_{[t_0,\tau]},y_{[t_0,\tau]}] \nonumber\\
&\qquad\times\ln{\frac{p[x_{[t_0,\tau]},y_{[t_0,\tau]}]}{p^\dagger[x_{[\tau,t_0]},y_{[\tau,t_0]}]}}\geq 0.
\end{align}
Next, however, we consider the following quantity
\begin{align}
&\langle \exp[-(\Delta \mathcal{S}^{xy}_{\rm int}+\Delta \mathcal{S}_{\rm tot}^{x}+\Delta \mathcal{S}_{\rm tot}^{y}-\Delta \mathcal{S}_{\rm storage}^{x}]\rangle =\nonumber\\
&\quad\int dx_{[t_0,\tau]}\int dy_{[t_0,\tau]}\;p(x_{t_0},y_{t_0})p[x_{(t_0,\tau]},y_{(t_0,\tau]}|x_{t_0},y_{t_0}]\nonumber\\
&\quad\times\frac{p(x_{\tau},y_{\tau})p^\dagger[x_{(\tau,t_0]},y_{(\tau,t_0]}|x_{\tau},y_{\tau}]}{p(x_{t_0},y_{t_0})p[x_{(t_0,\tau]},y_{(t_0,\tau]}|x_{t_0},y_{t_0}]}\frac{p[x_{[t_0,\tau]}]}{p^\dagger[x_{[\tau,t_0]}]}\nonumber\\
&=\int dx_{[t_0,\tau]}\;p^\dagger[x_{[\tau,t_0]}]\frac{p[x_{[t_0,\tau]}]}{p^\dagger[x_{[\tau,t_0]}]}\nonumber\\
&=1,
\end{align}
such that we have the bound
\begin{align}
\langle \Delta \mathcal{S}_{\rm tot}^{xy}-\Delta \mathcal{S}_{\rm storage}^{x}\rangle \geq 0
\label{bound3}
\end{align}
which may be considered equivalent to
\begin{align}
\langle\Delta\mathcal{S}_{\rm int}^{xy}\rangle+\langle\Delta \mathcal{S}_{\rm transfer}^{x\leftarrow y}\rangle\geq-\langle\Delta \mathcal{S}_{\rm tot}^y\rangle.
\end{align}
This, along with the property $\langle\Delta\mathcal{S}_{\rm storage}^{y}\rangle \geq 0$ indicates that if the total entropy production of subsystem $Y$ is negative, it arises because of \emph{negative} $\langle \Delta\mathcal{S}^{y\leftarrow x}_{\rm transfer}\rangle$ components that outweigh the information storage entropy, but moreover that this negative entropy is balanced by the entropy due to \emph{transfer} of information into $X$ along with the interaction entropy. In the bipartite systems considered here, in the steady state, this may be expressed
\begin{align}
\langle\Delta \mathcal{S}_{\rm transfer}^{x\leftarrow y}\rangle\geq-\langle\Delta \mathcal{S}_{\rm tot}^y\rangle
\end{align}
since the mutual information is unchanged, demonstrating explicitly how the extrinsic entropy production in $X$ controls the second law breakage in $Y$.
\par
Again, as with the treatment of switching protocols without feedback, we see a direct inverse relationship between the irreversibility in an extraneous time series, here $Y$, and the irreversibility in $X$ that arises from information transfer from that variable. Despite the presence of feedback, the mechanism is the same as in the case of a stochastic protocol. With increased irreversibility in $Y$, reflected in a positive $\Delta \mathcal{S}_{\rm tot}^y$, most sequences $y_{[t_0,\tau]}$ are common trajectories and thus well reflected in the coarse grained dynamics making them not particularly informative, whilst $y_{[\tau,t_0]}$ are rare and are thus more informative leading to negative entropy production due to information transfer. On the other hand when $\Delta \mathcal{S}_{\rm tot}^y$ is negative, sequences $y_{[t_0,\tau]}$ are chosen that are relatively uncommon (with respect to their time reverse, $y_{[t_0,\tau]}$, in construction of the marginal dynamics), leading to more informative realisations in the dynamics of $Y$ in the forward time direction and thus a positive entropy production due to information transfer.
\subsubsection{Recovery and generalization of previous information-thermodynamic bounds}
\label{TEbound}
Here we relate the bounds that we have derived, based on describing the thermodynamic entropy production as formed from information processing quantities, with previous results in the literature which generally seek to sharpen or generalize the second law with information processing quantities.
\par
A well known result concerning the transfer entropy rate in the steady state found in \cite{ito_maxwells_2015,horowitz_secondlawlike_2014,hartich_stochastic_2014} reads
\begin{align}
\langle\Delta\dot{\mathcal{S}}_{\rm tot}^x\rangle+\dot{T}_{x\to y}\geq 0
\end{align}
which may be integrated over finite times by considering the pathwise transfer entropy \cite{spinney_transfer_2017}
\begin{align}
\mathcal{T}_{x\to y}[x_{[t_0,t]},y_{[t_0,t]}]&=\ln\frac{p[y_{(t_0,t]}|y_{t_0},\{x_{[t_0,t)}\}]}{p[y_{(t_0,t]}|y_{t_0}]}
\end{align}
such that
\begin{align}
\langle\Delta{\mathcal{S}}_{\rm tot}^x\rangle+\langle\mathcal{T}_{x\to y}\rangle\geq 0.
\end{align}
Outside of the steady state an analogous expression for Bayesian networks was given in \cite{ito_information_2013} by considering the addition of a mutual information term, which after adapting to continuous time processes yields
\begin{align}
\langle\Delta{\mathcal{S}}_{\rm tot}^x\rangle+\langle\mathcal{T}_{x\to y}\rangle-\Delta I_M\geq 0.
\end{align}
We emphasize, the distinction between this relation and the analogous relation in our framework
\begin{align}
\langle\Delta{\mathcal{S}}_{\rm tot}^x\rangle+\langle\Delta\mathcal{S}_{\rm transfer}^{y\leftarrow x}\rangle+\langle\Delta\mathcal{S}_{\rm int}^{xy}\rangle\geq 0,
\end{align}
is that all terms in the latter are understood as components of the total entropy production of the composite system by considering the decomposition of the observed irreversibilities along computational lines. In this way, $\langle\Delta\mathcal{S}_{\rm int}^{xy}\rangle=-\Delta I_M$ is not an ad-hoc introduction to satisfy the bound, but the irreversibility associated with the joint prediction/computation in the system which must be accounted for. In contrast $\langle\Delta\mathcal{S}_{\rm transfer}^{y\leftarrow x}\rangle\neq \langle\mathcal{T}_{x\to y}\rangle$, with the former identified as a component of the entropy production in $Y$ whilst the latter is a pure information term. However, our formalism can be seen to reduce to the above if a specific choice of reverse process is considered, common to the original literature on feedback protocols \cite{sagawa_nonequilibrium_2012}, whereby the reverse process forces the extraneous variable, e.g. $Y$, to exactly retrace the path it traced in the forward realisation, independently of $X$ such that the reverse dynamics are distinct from the forward dynamics, i.e. $p\neq p^\dagger$. This has the specific result of setting $p^\dagger[y_{[\tau,t_0]},\{x_{[\tau,t_0]}\}]=p[y_{[t_0,\tau]}]=p^\dagger[y_{[\tau,t_0]}]$, precisely returning $\langle\Delta\mathcal{S}_{\rm transfer}^{y\leftarrow x}\rangle= \langle\mathcal{T}_{x\to y}\rangle$.
\par
This may give some insight into the thermodynamic meaning, if any, of the bare transfer entropy as it is equal to the \emph{irreversibility} $\langle\Delta \mathcal{I}^T_{y\leftarrow x}\rangle$ analogous to the the entropy production $\langle\Delta\mathcal{S}_{\rm transfer}^{y\leftarrow x}\rangle$ in the case of non-autonomous feedback. I.e. the transfer entropy is a limit case of  $\langle\Delta\mathcal{S}_{\rm transfer}^{y\leftarrow x}\rangle$ when $Y$ has no direct energetic, or thermal, interpretation as in the case of a non-autonomous feedback controller.
\par
There is no general bound characterizing the relationship between $\langle\Delta\mathcal{S}_{\rm transfer}^{y\leftarrow x}\rangle$ and $\langle\mathcal{T}_{x\to y}\rangle$, with the former in certain situations being larger or smaller than the transfer entropy depending on the nature of the dynamics and the reverse behaviour. However, their equality in the case of non-autonomous protocols, where it is relevant, allows confirmation of the bound in cases where $\langle\mathcal{T}_{x\to y}\rangle$ is computable, but $\langle\Delta\mathcal{S}_{\rm transfer}^{y\leftarrow x}\rangle$ is not, which we shall find of use when we illustrate the generalised application of our results in a non-bipartite system consisting of a thermodynamic system of interest and a feedback controller.
\section{Beyond the bipartite assumption}
\label{nonbip}
In this section we apply our formalism to systems which do not possess the bipartite structure considered thus far. Whilst the thermodynamics and entropy production of such systems may, at least in the case of discrete state spaces, be treated as a whole using conventional stochastic thermodynamics, treatment of the entropy production associated with their individual components ordinarily cannot. As discussed, this inability arises because under such dynamics the heat exported to the environment does not divide into components that can be exclusively associated with each sub-system. Rather, at least in some sense, there must be a component of the heat flow which is shared between the subsystems. When considering only discrete space, continuous time systems such as those described as a master equation it might seem natural to categorize such heat flows based on the nature of the transition, with a shared heat flow being associated with those not permitted in bipartite systems, namely joint transitions. However, in effect, this is merely categorizing distinct transitions as bipartite (individual) or non-bipartite (joint) and leaves fundamental questions unanswered. For instance, if \emph{all} transitions are joint transitions one would conclude that all heat flow should be identified as belonging to the joint system. However, one may still track only the probabilistic behaviour of $X$ in the context of $Y$ and seek to characterise its irreversibility. Is this irreversibility related to the irreversibility of the joint system or that of sub-system $Y$? What if only the energetics of one sub-system can be defined if the other corresponds to a non-thermal process such as a feedback controller? These questions go entirely unanswered if the formalism simply assigns all irreversibility to the composite system. Furthermore, in the limit of continuous dynamics the bipartite condition interpreted as non-coincident transitions is no longer applicable with such a property resulting in conditional independence in the dynamics. In these systems both $X$ and $Y$ simultaneously fluctuate on all timescales and so the identification of heat flows is generally non-trivial, but not necessarily ill defined. For instance, as we shall explore later, such continuous, non-bipartite, systems, if representing a physical system and correlated feedback controller, the heat dissipated by the physical system can be captured by the computational irreversibility.
\par
It is in answering these questions that we propose our formalism has particular merit since, because it is based on information theory, it is agnostic to the precise nature of the physical dynamics which are leading to the probability distributions, yet is consistent with conventional stochastic thermodynamics where heat flows can be identified as per the previous section.
\par
Explicitly we concern ourselves with systems where the relation in Eq.~(\ref{bipbip}) does not hold, i.e.
\begin{align}
&\lim_{dt\to 0}p(x_{t+dt},y_{t+dt}|x_t,y_t)\nonumber\\
&\qquad\neq \lim_{dt\to 0}p(x_{t+dt}|x_t,y_t)p(y_{t+dt}|x_t,y_t).
\end{align}
 To proceed, however, we take precisely the same course of action as before, establishing the interaction irreversibility based on a principle of conservation of predictive capacity according to Eq.~(\ref{cint}), which must hold independently of the physical energetics of the system or whether they are identifiable. This is given by
\begin{align}
&\Delta \mathcal{I}^{\rm int}_{xy}=-\Delta I_M\nonumber\\
&+\ln\frac{p[x_{(t_0,\tau]},y_{(t_0,\tau]}|x_{t_0},y_{t_0}]}{p^\dagger[x_{(\tau,t_0]},y_{(\tau,t_0]}|x_{t},y_{t}]}\nonumber\\
&+\ln\frac{p^\dagger[x_{(\tau,t_0]}|x_{t},\{y_{[\tau,t_0)}\}]p^\dagger[y_{(\tau,t_0]}|y_{t},\{x_{[\tau,t_0)}\}]}{p[x_{(t_0,\tau]}|x_{t_0},\{y_{[t_0,\tau)}\}]p[y_{(t_0,\tau]}|y_{t_0},\{x_{[t_0,\tau)}\}]}\nonumber\\
&=-\Delta I_M+\beta\Delta Q_{\rm int}^{xy}.
\label{DSint}
\end{align}
Explicitly, the interaction irreversibility is comprised of a change in mutual information which appeared in the bipartite description \emph{and} a heat contribution that did not appear. Again, this heat is \emph{not} identified by joint transitions in the composite system, but rather by the irreversibility which cannot be observed in the individual subsystems. At this point we remain agnostic as to whether $\Delta\mathcal{S}_{\rm tot}^x$ has a precise meaning in these circumstances (though we shall argue that it can in a subsequent example) and so strictly we may deal with the computation irreversibilities $\Delta\mathcal{I}_{x}=\Delta \mathcal{I}_x^A+\Delta \mathcal{I}_{x\leftarrow y}^T$. However,  the total entropy production of the composite system is well defined and so we may still identify the contributions to this composite entropy production in terms of storage and transfer, i.e. $\Delta\mathcal{I}_{x}=\Delta\mathcal{S}_{\rm storage}^x+\Delta\mathcal{S}_{\rm transfer}^{x\leftarrow y}$ and $\Delta \mathcal{I}_{\rm int}^{xy}=\Delta \mathcal{S}_{\rm int}^{xy}$, such that
\begin{align}
\Delta \mathcal{S}_{\rm tot}^{xy}&=\Delta \mathcal{I}_{x}+\Delta \mathcal{I}_{y}+\Delta \mathcal{I}^{\rm int}_{xy}\nonumber\\
&=\Delta \mathcal{S}_{\rm storage}^{x}+\Delta \mathcal{S}_{\rm storage}^{y}\nonumber\\
&+\Delta \mathcal{S}_{\rm transfer}^{x\leftarrow y}+\Delta \mathcal{S}_{\rm transfer}^{y\leftarrow x}+\Delta\mathcal{S}_{\rm int}^{xy}\nonumber\\
&=\Delta\mathcal{S}_{\rm sys}^x+\beta\Delta Q^x_{\rm storage}+\Delta\mathcal{S}_{\rm sys}^y+\beta\Delta Q^y_{\rm storage}\nonumber\\
&+\beta\Delta Q^{x\leftarrow y}_{\rm transfer}+\beta\Delta Q^{y\leftarrow x}_{\rm transfer}-\Delta I_M+\beta\Delta Q^{xy}_{\rm int}
\end{align}
such that
\begin{align}
\Delta Q^{xy}&=\Delta Q^x_{\rm storage}+\Delta Q^y_{\rm storage}+\Delta Q^{x\leftarrow y}_{\rm transfer}+\Delta Q^{y\leftarrow x}_{\rm transfer}\nonumber\\
&+\Delta Q^{xy}_{\rm int}.
\end{align}
At this point we reiterate, these are divisions of the total heat which is well defined, but due to the lack of bipartite dynamics cannot necessarily associate $\Delta Q^x_{\rm storage}+\Delta Q^{x\leftarrow y}_{\rm transfer}=\Delta Q^x$ as we could in the previous section.
\par
Importantly, however, these entropy productions follow \emph{exactly} the same relationships as for the bipartite case, by construction, with the same integral fluctuation theorems holding, namely,
\begin{align}
\langle\exp[-(\Delta\mathcal{S}^{xy}_{\rm int}+\Delta\mathcal{S}^{x}_{\rm storage}+\Delta\mathcal{S}^{x\leftarrow y}_{\rm transfer}+\Delta\mathcal{S}^{y\leftarrow x}_{\rm transfer})]\rangle=1
\end{align}
again such that
\begin{align}
&\langle\Delta\mathcal{S}^{xy}_{\rm int}\rangle+\langle\Delta\mathcal{S}^{y\leftarrow x}_{\rm transfer}\rangle\nonumber\\
&\qquad\geq-\langle\Delta\mathcal{S}^{x}_{\rm storage}\rangle-\langle\Delta\mathcal{S}^{x\leftarrow y}_{\rm transfer}\rangle= -\langle\Delta\mathcal{I}^{x}\rangle,
\end{align}
as before, lending weight to the generality of the proposed measures of irreversibility. 
\par
Moreover, the above bound also holds when $\Delta\mathcal{S}^{y\leftarrow x}_{\rm transfer}$ is replaced by the bare pathwise transfer entropy $\mathcal{T}_{x\to y}$, as in the bipartite case, allowing us to write 
\begin{align}
\langle\Delta\mathcal{I}_{x}\rangle+\langle\Delta\mathcal{I}_{xy}^{\rm int}\rangle\geq -\langle\mathcal{T}_{x\to y}\rangle
\end{align}
or rather
\begin{align}
\langle\Delta\mathcal{I}_{xy}\rangle-\langle\Delta\mathcal{I}_{y}\rangle\geq -\langle\mathcal{T}_{x\to y}\rangle
\end{align}
which may equivalently be expressed in the rate form $\langle\Delta\dot{\mathcal{I}}_{x}\rangle+\langle\Delta\dot{\mathcal{I}}_{xy}^{\rm int}\rangle\geq -\dot{T}_{x\to y}$. This is a central result and can be seen to generalize the well known information-thermodynamic bound $\langle\Delta\dot{\mathcal{S}}^x_{\rm tot}\rangle\geq -\dot{T}_{x\to y}$ and $\langle\Delta\dot{\mathcal{S}}^x_{\rm tot}\rangle\geq \Delta \dot{I}_M-\dot{T}_{x\to y}$ explored in depth in several other works \cite{horowitz_secondlawlike_2014,ito_information_2013,ito_maxwells_2015,hartich_stochastic_2014,hartich_sensory_2016} to the case of non-bipartite dynamics. Explicitly, when bipartite dynamics are employed $\Delta\dot{\mathcal{I}}_{x}$ reduces to $\Delta\dot{\mathcal{S}}^{x}_{\rm tot}$ and $\Delta\dot{\mathcal{I}}_{xy}^{\rm int}$ reduces to $\Delta\dot{I}_{M}$ which then vanishes in the steady state. It is important to note that in both regimes the transfer entropy $\dot{T}_{x\to y}$ bounds the irreversibility associated with the subsystem $X$ \emph{and} the irreversibility associated with its interaction with $Y$.
\par
It is interesting to consider what physical systems might be described by such dynamics that do not permit bipartite form. Two major examples can be considered. The first is when the whole system is well defined thermodynamically, with defined heat flows and dissipation, but has interacting parts that evolve in such a way that it cannot be traditionally decomposed any further. Alternatively when considering autonomous measurement and feedback one can imagine a feedback mechanism which does not meet the bipartite assumption. In both cases either the total heat of the composite system or the the heat flow in system being monitored should still be identifiable. However, currently no formalism exists for characterizing any such division of this total heat or for characterizing the feedback performance in these situations. A particular case will be treated in the next section. 
\section{Examples}
\subsection{Minimal bipartite model}
Here we illustrate some of the behaviour of the entropy production contributions with a simple model. This model consists of two subsystems interacting with a bipartite structure. The first subsystem, $X$, is comprised of three states $x\in\mathcal{X}=\{A,B_1,B_2\}$, whilst the second subsystem, $Y$, is comprised of two states, $y\in\mathcal{Y}=\{1,2\}$ such that there are $6$ effective global states $\{x,y\}\in \mathcal{X}\otimes\mathcal{Y}$. The bipartite structure insists upon strict conditional independence in the updating of these subsystems, which, in continuous time, means that there are no instantaneous joint transitions in both $X$ and $Y$ with probability $1$.
\par
The structure is an elaboration on the minimal model introduced in \cite{hartich_stochastic_2014}, but with an additional state in subsystem $X$. Further, we insist upon a particular transition rate structure which entails the transition rates governing subsystem $Y$ only being dependent on whether $X$ is in state $A$ or the collection of states $B=\{B_1,B_2\}$, such that it cannot distinguish between states $B_1$ and $B_2$ that comprise $B$. This structure is illustrated in Fig.~(\ref{fig2}) along with specific transition rates, specified up to the transition rates between the sub-states $B_1$ and $B_2$. Other transitions are parametrized by the dimensionless constants $\{r_x,r_y\}\in[-1,1]\otimes[-1,1]$ and timescale parameters $\gamma_x>0$, $\gamma_y>0$.
\par
This structure allows us to separate the mechanisms that pertain to irreversibility associated with storage  and transfer of information. This is because the transitions between $B_1$ and $B_2$ can introduce irreversibility independently of $Y$, whilst the transitions between the pairs $A$, $B$ and $1$, $2$ explicitly interact with each other, but do not introduce any net steady current into either subsystem when viewed in isolation.
\par
To this end we consider a parameterization of the transition rates $\kappa_y^{B_1\leftrightarrow B_2}$ as follows
\begin{align}
\kappa_y^{B_1\rightarrow B_2}&=\gamma_x k^+, \quad \forall y\in\{1,2\}\nonumber\\
\kappa_y^{B_2\rightarrow B_1}&=\gamma_x k^-, \quad \forall y\in\{1,2\}
\end{align}
where $k^+$ and $k^-$ are arbitrary constants in $(0,1)$. Considering the steady state and solving the master equation by standard methods then yields 
\par
\begin{align}
\langle\Delta\dot{S}_{\rm storage}^x\rangle&=\frac{\gamma_x(r_x+r_y-2r_xr_y)(k^+-k^-)\ln[k^+/k^-]}{3\left((k^++k^-)+(r_x+r_y-2r_xr_y)\right)}\nonumber\\&\quad+\mathcal{O}(a)\nonumber\\
\langle\Delta\dot{S}_{\rm transfer}^{x\leftarrow y}\rangle&=\frac{4}{3}\gamma_x(r_x-r_y)\ln\left[\frac{1-r_x}{r_x}\right]+\mathcal{O}(a)\nonumber\\
\langle\Delta\dot{S}_{\rm transfer}^{y\leftarrow x}\rangle&=\frac{4}{3}\gamma_x(r_y-r_x)\ln\left[\frac{1-r_y}{r_y}\right]+\mathcal{O}(a)\nonumber\\
\langle\Delta\dot{S}_{\rm storage}^y\rangle&=0
\label{32res}
\end{align}
where $a=\gamma_x/\gamma_y$ such that we have effected a separation of timescales in order to avoid the difficulty associated with calculation of generally non-Markovian objects. If a further separation of timescales exists, $\kappa_y^{B_1\rightarrow B_2}+\kappa_y^{B_2\rightarrow B_1}=\gamma_x'\ll\gamma_x$ then the storage entropy reduces to $\langle\Delta\dot{S}_{\rm storage}^x\rangle=(1/3)\gamma_x'(k^+-k^-)\ln[k^+/k^-]+\mathcal{O}(\gamma_x'/\gamma_x)$.
\begin{figure*}[!htb]
\includegraphics[width=0.5\textwidth]{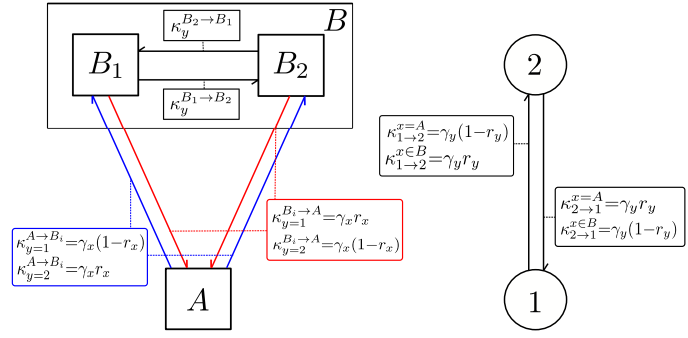}
\caption{\label{fig2}Coupled three and two state systems. System X (shown on left) has transition rates dependent on the instantaneous state of system Y. System Y (shown on right) has transition rates dependent on the instantaneous state of system X. System Y can only depend on the coarse states $A$ and $B$ in X: the transition rates of system Y \emph{do not} distinguish between states $B_1$ and $B_2$. Since the system is bipartite, no joint transitions in systems X and Y can occur. }
\end{figure*}
The results in Eqs.~(\ref{32res}) are illustrated in Fig.~(\ref{fig3}). Importantly we see that as we vary the feedback coefficient $r_y$ controlling subsystem $Y$ the entropy due to transfer in each system is offsetting the other, with the entropy production associated with storage only varying weakly based on changes in the stationary solution to the master equation. Indeed, since in this example we have $\langle\Delta\dot{\mathcal{S}}^{y}_{\rm storage}\rangle=0$, a stronger bound than Eq.~(\ref{bound3}) holds, with $\langle\Delta\dot{\mathcal{S}}^{x\leftarrow y}_{\rm transfer}+\Delta\dot{\mathcal{S}}^{y\leftarrow x}_{\rm transfer}\rangle\geq 0$ holding completely independently of $\langle\Delta\dot{\mathcal{S}}^{x}_{\rm storage}\rangle$ in addition to $\langle\Delta\dot{\mathcal{S}}^{x}_{\rm tot}+\Delta\dot{\mathcal{S}}^{y\leftarrow x}_{\rm transfer}\rangle\geq 0$. Moreover, if we consider $r_x$ and $r_y$ to be fixed, varying only the thermodynamic force causing asymmetry in the $B_1\leftrightarrow B_2$ transition, we see total independence of in the entropy productions due to transfer as the asymmetric transitions are not implicated in the feedback between the two systems. Indeed we see that the entropy production due to storage in this example manifests exclusively due to stationary current which can only arise when $k^+$ and $k^-$ are unequal explaining the behaviour with varying $k^+$.
\begin{figure*}[!htb]
\includegraphics[width=0.4\textwidth]{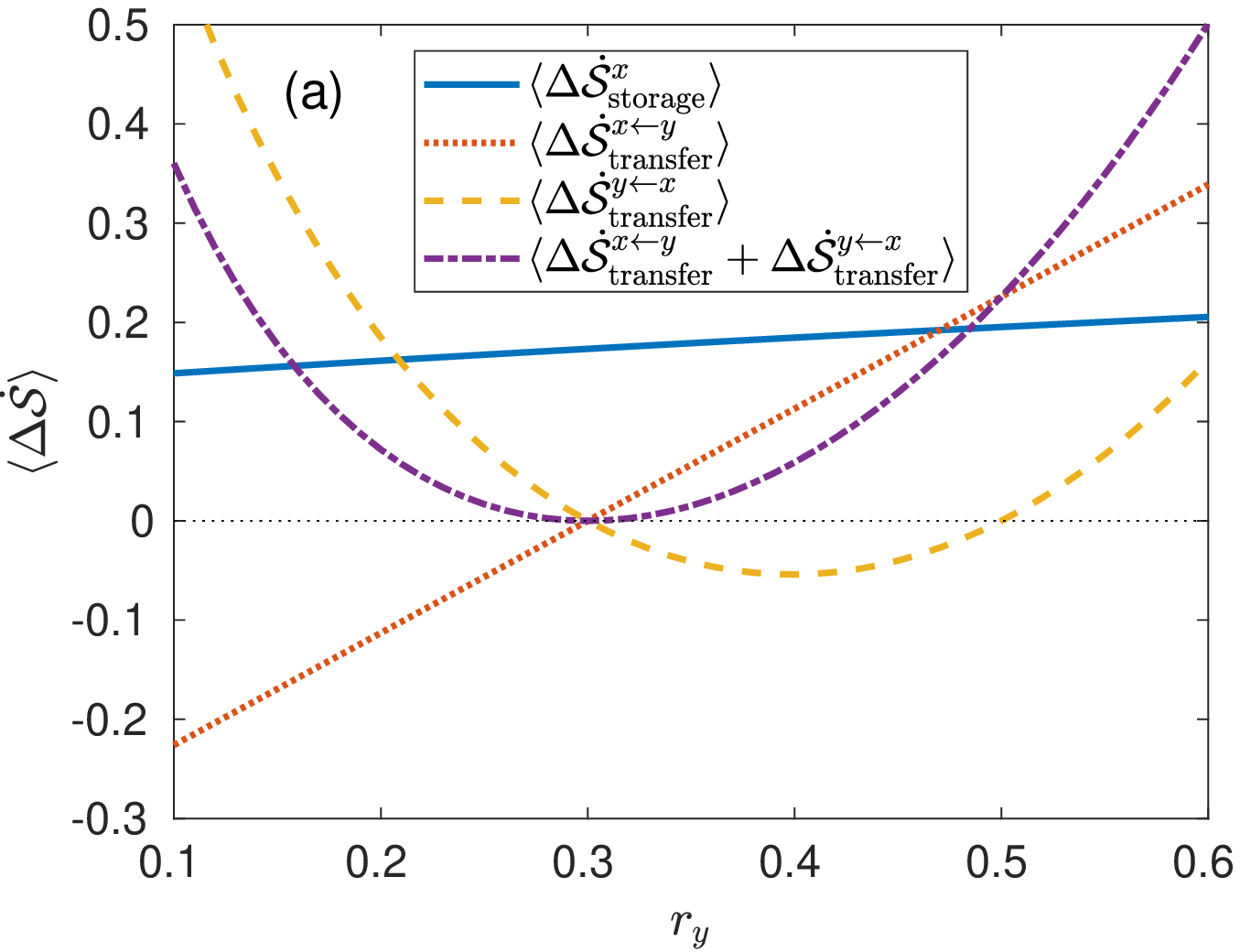}
\includegraphics[width=0.4\textwidth]{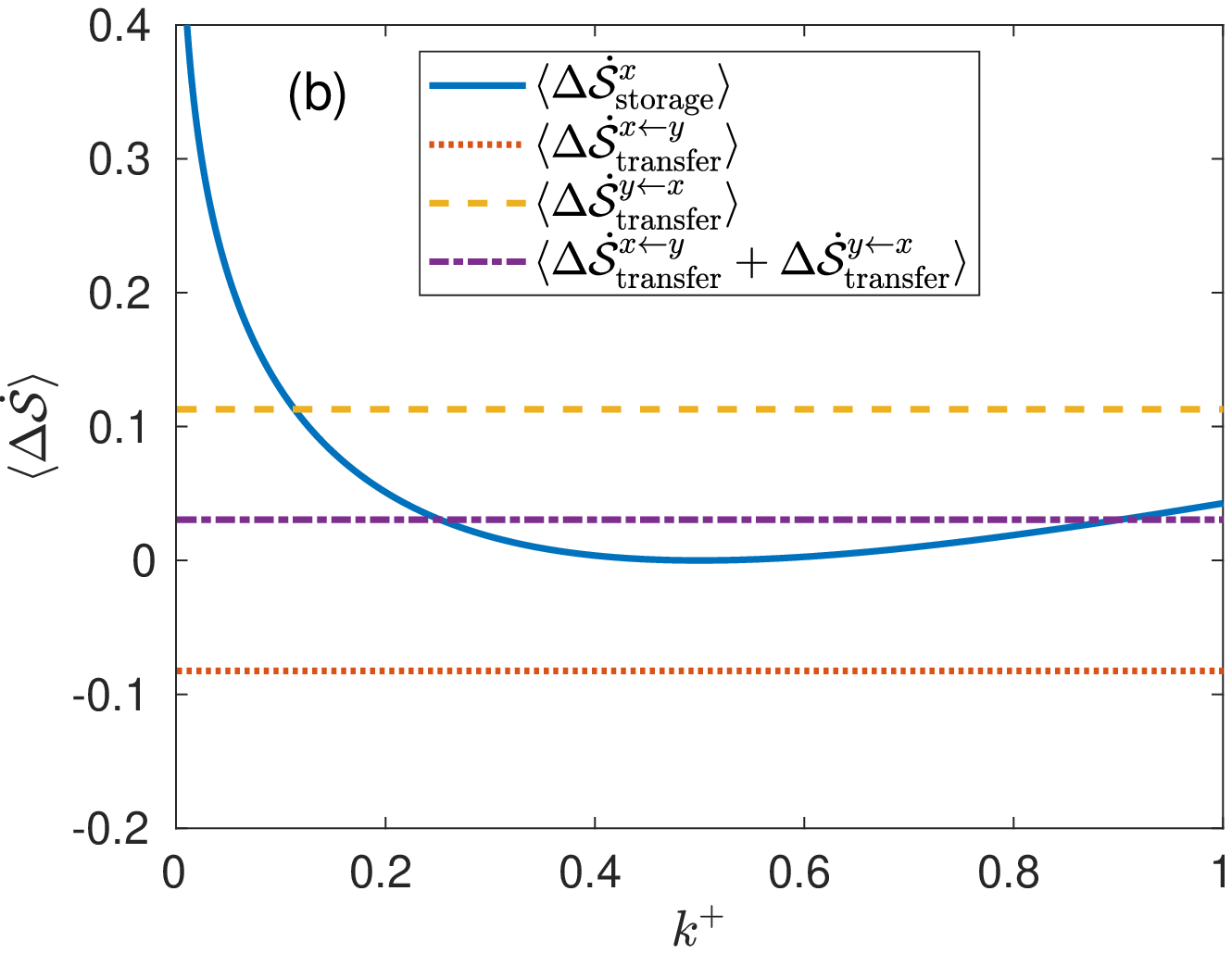}
\caption{\label{fig3}Entropy productions arising from the coupled three and two state model. In sub-figure (a) we have set $\gamma_x=1$, $k^-=0.1$, $k^+=0.9$, $r_x=0.3$ and allow $r_y$ to vary. In sub-figure (b) we have set $\gamma_x=1$, $k^-=0.5$, $r_x=0.3$, $r_y=0.3$ and allow $k^+$ to vary. Both consider $0$-th order contributions in $a=\gamma_x/\gamma_y$.}
\end{figure*}
\subsection{Computational and physical irreversibility in non-bipartite autonomous feedback systems}
\label{nonbipsec}
Generally, there are many systems that do not obey the bipartite assumption. These range from composite systems which permit joint transitions between different elements, composite systems where the subsystems experience thermal (or other) noise sources that cannot be reliably idealised as being independent through to systems traditionally out of scope thermodynamics. However, an investigation of their irreversibility may yet yield analogous bounds important to their behaviour.
\par
In this section, however, we consider a particular example, that is a system designed to model autonomous feedback by a controller, $Y$, on a physical system of interest $X$. A minimal mathematical model that achieves this is a pair of coupled Ornstein Uhlenbeck processes described by the stochastic differential equations
\begin{align}
dx&=Axdt+Bydt+V_xdW_x\nonumber\\
dy&=Cxdt+Dydt+V_ydW_y,
\label{sdes}
\end{align}
where $dW_x$ and $dW_y$ represent thermal noise and the inherent error in the measurement process respectively. Normally these two noise sources are deemed to be statistically independent resulting in bipartite dynamics, the behaviour of which has been previously studied \cite{ito_maxwells_2015,allahverdyan_thermodynamic_2009,horowitz_secondlawlike_2014}. However, we can quite easily posit that the noise sources \emph{are} correlated for any number of plausible physical reasons, for example noise in the measurement apparatus may be arising, in part, from the same thermal source such that the measurement suffers from noise contamination. In this case we have $\langle dW_x dW_y\rangle=\rho dt$ where $\rho\in[-1,1]$ characterises the noise correlation such that for $\rho=1$ the noise experienced by each of $X$ and $Y$ are identical.
\par
This system is governed by a Fokker-Planck equation of the form
\begin{align}
\partial_t p(x,y,t)&=-\nabla\cdot(J_x,J_y)\nonumber\\
J_x&=(Ax+By)p(x,y)-\frac{V_x^2}{2}\frac{\partial p(x,y)}{\partial x}\nonumber\\
&\qquad-\frac{\rho V_xV_y}{2}\frac{\partial p(x,y)}{\partial y}\nonumber\\
J_y&=(Cx+Dy)p(x,y)-\frac{V_y^2}{2}\frac{\partial p(x,y)}{\partial y}\nonumber\\
&\qquad-\frac{\rho V_xV_y}{2}\frac{\partial p(x,y)}{\partial x}
\label{FP1}
\end{align}
which has a Gaussian steady state given in Appendix \ref{appAA}.
\par
A well known bound that arises in information thermodynamics is  \cite{ito_maxwells_2015,horowitz_secondlawlike_2014,hartich_stochastic_2014}, as usually stated in the steady state,
\begin{align}
\frac{d\beta\langle\Delta Q^x\rangle}{dt}+\dot{T}_{x\to y}&\geq 0
\end{align}
where 
$\dot{T}_{x\to y}$ is transfer entropy rate. We emphasize, with correlated noise this relation \emph{does not} hold. 
\subsubsection{Identification of information and heat flows}
Whilst the framework outlined in section \ref{nonbip} describes how to divide entropy productions along computational divisions in terms of predictive capacity to arbitrary systems where individual currents and heats may not be well defined, we emphasize that for continuous systems of the form in Eq.~(\ref{sdes}) we \emph{can} identify such quantities, under the right circumstances, despite the lack of a bipartite structure.
\par
To do so we point out that the sampling paths are absolutely continuous such that, unlike with master equations with joint transitions, the variables never change discontinuously. As such the notion of a dissipated heat, by subsystem $X$, for instance, is not inappropriate, but more challenging to identify and depends on the exact nature of the model. For instance in the present model it is of consequence that $X$ is the thermodynamic system in question whilst $Y$ represents a feedback controller. As such we are not concerned with the thermodynamics of $Y$ or even the joint system $\{X,Y\}$ and this changes the interpretation. As such, the entropy production of the dynamics of the joint particle and controller system is immaterial, but their computational irreversibilities still are well defined. 
\par
To proceed and identify such heat flows associated with the single subsystem $X$, we must be more careful than just assuming a conventional local detailed balance relation, but rather appeal to the arguments in the spirit of the foundational stochastic energetics of Sekimoto \cite{ken_sekimoto_stochastic_2010,sekimoto_langevin_1998,sekimoto_kinetic_1997}. First we consider systems, such as over-damped Langevin dynamics where the energetics are all determined by the random variables such that all differentials are stochastic too. In such situations we state that there is some Hamiltonian, $H$, controlling $X$ which is parametrized by a protocol $Y$. When this protocol is deterministic, this traditionally leads to a stochastic version of the first law as follows
\begin{align}
dH(x,y)&=\partial_x H(x,y)\circ dx+\partial_y H(x,y)\dot{y}dt,
\end{align}
identifying changes in internal energy, heat and work ($dU=dQ+dW$) with the crucial use of Stratonovich integration rules (indicated by the $\circ$ notation) which render the stochastic calculus in the form of the usual thermodynamic differentials. If we allow the protocol $Y$ to be stochastic, we instead write
\begin{align}
dH(x,y)&=\partial_x H(x,y)\circ dx+\partial_y H(x,y)\circ dy
\label{FL2}
\end{align}
with the division of internal energy into heat and work operating in exactly the same way. 
Moreover, Eq.~(\ref{FL2}) remains unchanged under correlated noise such that the heat and work contributions are still identifiable where we note that Stratonovich integration includes correlation terms in the non-bipartite case such that
\begin{align}
f(x,y)\circ dx&=f(x,y)dx+\frac{V_x^2}{2}\frac{\partial f(x,y)}{\partial x}dt\nonumber\\
&\quad+\frac{\rho V_xV_y}{2}\frac{\partial f(x,y)}{\partial y}dt
\end{align} 
where absence of the $\circ$ notation indicates the non-anticipating It\={o} integral. Under this interpretation we might posit some over-damped dynamics such that $X$ is a spatial coordinate, $V_x=\sqrt{2k_BT/m\gamma}$ and $Ax+By=-(m\gamma)^{-1}\partial_xH_x(x,y)$ such that we have a Hamiltonian $H_x(x,y)=-m\gamma((A+B)x^2-B(x-y)^2)$, i.e. a harmonic trap with a protocol controlling an additional harmonic term. Importantly, in this system the computational irreversibility reflects the ratio of heat flow to the environment to the environmental temperature, $\beta\partial_x H(x,y)\circ dx$, such that $\Delta\mathcal{I}_x=\Delta\mathcal{S}^{x}_{\rm storage}+\Delta\mathcal{S}^{x\leftarrow y}_{\rm transfer} =\Delta\mathcal{S}^{x}_{\rm sys}+\beta\Delta Q^x$. 
\par
On the other hand, in situations where the system contains degrees of freedom which are not stochastic, but appear in the Hamiltonian an additional correlation term appears in the computational irreversibility which has no thermodynamic analog. In these cases we must instead identify
\begin{align}
\Delta\mathcal{I}_x=\Delta\mathcal{S}_{\rm sys}^x+\beta\Delta Q^x+\Delta\mathcal{S}_{\rm corr}^x
\end{align}
where $\Delta\mathcal{S}_{\rm corr}^x$ is the difference in heat flows between that which can be physically associated with $X$ and that which is associated with the irreversibility of $X$ such that
\begin{align}
d\mathcal{S}_{\rm corr}^x&=\beta dQ^x_{\rm corr}=\ln\frac{p^\dagger(x_{t}|x_{t+dt},y_{t+dt})}{p^\dagger(x_{t}|x_{t+dt},y_{t})}.
\end{align}
This situation arises due to parity differences in the variables separating the current into reversible and irreversible components \cite{spinney_entropy_2012}. Consequently differential forms for the environmental entropy production introduce a correlation term that would arise from stochastic integration of the reversible current, but does not emerge in the stochastic energetics since the energetic term arises from integration with respect to some deterministic variable (i.e. one not subject to noise), e.g. the particle position in a full under-damped Langevin description. We emphasize, in such situations great care must be taken to identify the exact nature of the dynamics and any Hamiltonian, if any, they are emerging from in order to correctly identify irreversible and reversible currents. We also mention that correlated noise must be treated carefully in systems with odd variables where the noise itself may be required to transform under time reversal \cite{spinney_use_2012}.
\par
For the remainder we shall consider the case of the former, where, in the steady state we consider $\langle\Delta\mathcal{S}_{\rm storage}^x+\Delta\mathcal{S}_{\rm transfer}^{x\leftarrow y}\rangle=\beta \langle\Delta Q^x\rangle$, for brevity, but acknowledge the possibility of this additional step in the identification of the environmental entropy production in different models.
\par
Returning to the model in question we reiterate that the naive application of the traditional information theoretic bound on the computational/physical irreversibility does not hold, i.e.
\begin{align}
\frac{d\langle\Delta \mathcal{S}_{\rm storage}^x\rangle}{dt}+\frac{d\langle\Delta \mathcal{S}_{\rm transfer}^{x\leftarrow y}\rangle}{dt}+\dot{T}_{x\to y}&\ngeq 0.
\end{align}
Instead, in order to find reliable bounds we must turn to our formulation in terms of computational irreversibilities which accounts for the interaction entropy production. As such the appropriate bound is 
\begin{align}
\frac{d\langle\Delta \mathcal{S}_{\rm tot}^x\rangle}{dt}+\frac{d\langle\Delta \mathcal{I}^{\rm int}_{xy}\rangle}{dt}+\frac{d\langle\Delta \mathcal{S}_{\rm transfer}^{y\leftarrow x}\rangle}{dt}\geq 0
\end{align}
or
\begin{align}
\frac{d\langle\Delta \mathcal{S}_{\rm tot}^x\rangle}{dt}+\frac{d\langle\Delta \mathcal{I}^{\rm int}_{xy}\rangle}{dt}+\dot{T}_{x\to y}\geq 0.
\label{boundh}
\end{align}
to account for the non-bipartite dynamics. We note the inclusion of the more general $\Delta \mathcal{I}^{\rm int}_{xy}$ rather that $\Delta \mathcal{S}_{\rm int}^{xy}$ in the bound because, as discussed, whilst the computational irreversibility of the joint particle/controller system is well defined, it is not related to an entropy production in this case.
\par
For the present example we are in a position to illustrate such a bound mathematically. In the steady state, the transfer entropy of this system permits an analytical solution, adapted from \cite{horowitz_secondlawlike_2014} (see also \cite{barnett_detectability_2017}), derived in appendix \ref{appBB},
\begin{align}
\dot{T}_{x\to y}&=\frac{1}{2}\left(\sqrt{A^2-2\rho AC\frac{V_x}{V_y}+C^2\frac{V^2_x}{V^2_y}}-\sqrt{A^2}-\rho C\frac{V_x}{V_y}\right).
\end{align}
Meanwhile, the rate of computational irreversibility in $X$ in the steady state, here equal to the entropy production rate, is given by
\begin{align}
\frac{d\langle\Delta\mathcal{S}^x_{\rm tot}\rangle}{dt}&=\frac{B(CV_x^2-V_y(\rho(A-D) V_x +BV_y))}{(A+D)V_x^2}.
\end{align}
For completeness we also point out that for this system we have
\begin{align}
\frac{d\langle\Delta\mathcal{S}^x_{\rm corr}\rangle}{dt}&=\frac{\rho BV_y}{V_x}.
\end{align}
We may then in turn compute the interaction entropy rate, again in the steady state, which due to a slightly cumbersome form we report in appendix \ref{appAA}.
Together, these objects obey the bound in Eq.~(\ref{boundh}). These are illustrated in Fig.~(\ref{fig4}) where they are contrasted with the usual bipartite relation and, again for completeness, the bound with $\Delta\mathcal{S}^x_{\rm tot}$ replaced with $\Delta\mathcal{S}^x_{\rm tot}-\Delta\mathcal{S}_x^{\rm corr}$. 
Importantly we see that 
the previous conception of the generalised second law in such a system fails for some value of correlation $\rho$, 
holding for $\rho=0$ where the system reduces to the usual bipartite model.
\begin{figure*}[!htb]
\includegraphics[width=0.5\textwidth]{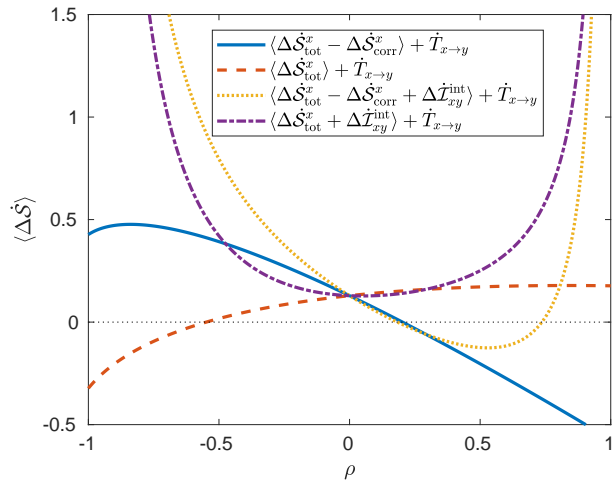}
\caption{\label{fig4}Coupled linear feedback model. We set $A=-2$, $D=-4$, $B=C=1$, $V_x=2$, $V_y=3/2$ and allow the correlation, $\rho$, to vary.}
\end{figure*}
\section{Discussion and Conclusions}
In this paper we have considered computations described in the framework of information dynamics. Using such a framework we have defined appropriate time reversed computations and shown that the difference between each results in a measure of irreversibility that can be divided into computationally meaningful quantities: those that relate to the storage and transfer of information. By applying such a framework to physical systems we have shown that such a measure of computational irreversibility maps onto the total entropy production in stochastic thermodynamics revealing  that not only can information processing lead to bounds on thermodynamics, as previously conceptualized, but that thermodynamics may be entirely described in terms of information processing.
\par
Moreover, the framework based on information dynamics that we have presented is not restricted to the bipartite and multipartite systems previously considered in stochastic thermodynamics. By dividing irreversibility along the lines of computational primitives, components of the total physical heat can be associated with irreversibility even when the physical heats exported by the sub-systems are ambiguous. Further we have shown that these irreversibilities based on storage and transfer are critically implicated in the resulting bounds that generalize the second law both when a bipartite assumption holds and when it does not. In both cases the bounds are achieved by correctly accounting for an interaction entropy production which is always involved in its formulation and derived from a principle of conserved predictive capacity. Further, through the use of continuous dynamics with feedback and correlated noise we have presented a system where the physical heat exported by a subsystem is identifiable despite being governed by non-bipartite dynamics. Importantly, for this system, we have demonstrated that previous generalizations of the second law which account for feedback fail, whilst the bound arising from this formalism holds.
\par
We hope that such a framework will allow discussion of irreversibility in systems where identification of individual heat flows are problematic and, moreover, allow discussion of thermodynamic processes in terms of distributed computation.
\begin{appendix}
\begin{widetext}
\section{Generalised computations}
Whilst not strictly necessary for the development of irreversibility measures, the intrinsic (forward) and time reversed computations can be unified through the notion of a generalised intrinsic computation. To define such an object we appreciate that the basic structure of the intrinsic computation is the reduction in uncertainty of the next step in a time series, relative to an ensemble prior using the past sequence of events and a transition probability. And it is this notion that we seem to generalize based on the recognition of its structure in terms of uncertainties. First we allow the prediction, or reduction of uncertainty, of some arbitrary event $\varepsilon$, taking values in some space $\mathcal{E}$, rather than merely the next step in the time series. This event must be computable with a sequence of states of variable $X$, that we write $\vec{X}$, taking values $\vec{x}\in\vec{\mathcal{X}}$. Consequently we write $\varepsilon=f(\vec{x})$ such that $f:\vec{\mathcal{X}}\to\mathcal{E}$. We then wish to discuss the realisations of $X$ that lead to the event $\varepsilon$ such that we consider the pre-image of $\varepsilon$
\begin{align}
f^{-1}(\varepsilon)=\{\vec{x}\in\vec{\mathcal{X}}|f(\vec{x})=\varepsilon\}.
\end{align}
Associated with this set of realisations $\vec{x}$ we have a prior distribution on $\vec{X}$, denoted, $\mathbb{Q}(\vec{x}\in \mathcal{A})$, which represents our prior uncertainty in sets of $\vec{x}$. As such, the prior uncertainty in the specific event $\varepsilon$ corresponds to $\mathcal{A}=f^{-1}(\varepsilon)$, i.e. $\mathbb{Q}(\vec{x}\in f^{-1}(\varepsilon))$.
\par
 Finally we consider an inference scheme that provides a posterior, which quantifies uncertainty in $\varepsilon$ in the presence of some evidence $\vec{Z}$ taking values $\vec{z}\in\vec{\mathcal{Z}}$, where, like $\vec{X}$, $\vec{Z}$ consists of some time series. This inference scheme is represented with a probability measure $\mathbb{P}(\vec{x}\in\mathcal{A}|\vec{z})$ where $\mathcal{A}\subseteq\vec{\mathcal{X}}$ where, again, the set corresponding to the event $\varepsilon$ is $\mathcal{A}=f^{-1}(\varepsilon)$, such that the uncertainty under inference scheme $\mathbb{P}$ and evidence $\vec{z}$ is $\mathbb{P}(\vec{x}\in f^{-1}(\varepsilon)|\vec{z})$.
\par
Specifying all such quantities defines a tuple $\mathcal{C}=\{\vec{X},f,\vec{Z},\mathbb{P},\mathbb{Q}\}$ that entirely characterises the local computational signature of this generalised computation through the consideration of
\begin{align}
c_\mathcal{C}&=\ln\frac{d\mathbb{P}(\vec{x}\in f^{-1}(\varepsilon)|\vec{z})}{d\mathbb{Q}(\vec{x}\in f^{-1}(\varepsilon))}.
\end{align}
In these terms we can identify the (forward) \emph{intrinsic computation}, at time $i$, as being characterized by the tuple
\begin{align}
& \mathcal{C}_{\{i,i\text{+}1\}}=\{X_{i\text{+}1:n},X_{i+1},\{X_{0:i},Y_{0:i}\},p_{X_{i\text{+}1:n}|X_{0:i},Y_{0:i}}(\cdot|\cdot),p_{X_{i\text{+}1:n}}(\cdot)\}.
 \end{align}
  I.e. the sequence of $X$ from which the event is determined is the future states of the system, $X_{i{+}1:n}$. The predicted event is the trivial choice of the next state the system evolves to, $\varepsilon=f(x_{{i{+}1:n}})={x}_{i{+}1}$. The pre-image $f^{-1}(\epsilon)$ is therefore the set of all paths $X_{i{+}1:n}$ that has $x_{i+1}=\varepsilon$, $f^{-1}(\varepsilon)=\{x_{i{+}1:n}\in\mathcal{X}_{i{+}1:n}:x_{i{+}1}=\varepsilon\}$. The evidence is the entirety of previous states of all available time series $\vec{Z}=\{X_{0:i},Y_{0:i}\}$. Using these objects the prior belief is then that of the inherent uncertainty in the future paths of the system  $\mathbb{Q}({x}_{i{+}1:n})=p(X_{i{+}1:n}=x_{i{+}1:n})\equiv p_{X_{i{+}1:n}}({x}_{i{+}1:n})$ and the inference strategy is provided by the same ensemble dynamics $\mathbb{P}(x_{i{+}1:n}|x_{0:i},y_{0:i})=p(X_{i{+}1:n}=x_{i{+}1:n}|X_{0:i}=x_{0:i},Y_{0:i}=y_{0:i})\equiv p_{X_{i\text{+}1:n}|X_{0:i},Y_{0:i}}(x_{i{+}1:n}|x_{0:i},y_{0:i}) $. One then returns the predictive capacity of the intrinsic computation by recognizing that
  \begin{align}
  p_{X_{i{+}1:n}|X_{0:i},Y_{0:i}}(x_{i{+}1:n}\in f^{-1}(x_{i{+}1}|x_{0:i},y_{0:i})&=  p_{X_{i{+}1}|X_{0:i},Y_{0:i}}(x_{i{+}1}|x_{0:i},y_{0:i})\nonumber\\
  &=  p_(X_{i{+}1}=x_{i{+}1}|X_{0:i}=x_{0:i},Y_{0:i}=y_{0:i})\nonumber\\
  p_{X_{i{+}1:n}}({x}_{i{+}1:n}\in f^{-1}(x_{i{+}1}))&=  p_{X_{i{+}1}}({x}_{i{+}1})=  p(X_{i{+}1}={x}_{i{+}1}).
  \end{align}
\par
In these terms one can define other information-theoretic quantities with other choices. For example, the predictive information, closely related to the excess entropy, results from the following choice, in the limit $n-i\to\infty$, $i\to \infty$
\begin{align}
& \mathcal{C}^{\rm pred}_{\{i,i\text{+}1\}}=\{X_{i\text{+}1:n},X_{i\text{+}1:n},X_{0:i},p_{X_{i\text{+}1:n}|X_{0:i}}(\cdot|\cdot),p_{X_{i\text{+}1:n}}(\cdot)\}.
 \end{align}
In contrast, the definition of the time reversed computation emerges through the tuple
\begin{align}
\mathcal{C}^\dagger_{\{i,i+1\}}=\{X_{i:0},X_i,\{X_{n:i\text{+}1},Y_{n:i\text{+}1}\},p^\dagger_{X_{n-i:n}|X_{0:n-i-1},Y_{0:n-i-1}}(\cdot,\cdot),p_{X_{i:0}}(\cdot)\}.
\end{align}
I.e. the sequence from which the predicted event is computed is the time reversed sequence of past states $\vec{X}=X_{i:0}$, the event is simply the `next' state in this sequence, $f(X_{i:0})=X_i$ and the evidence sequence is the collection of time reversed future states $\vec{Z}= \{X_{n:i\text{+}1},Y_{n:i\text{+}1}\}$. Using these objects the prior distribution $\mathbb{Q}_{\vec{X}}$ is, as before, the ensemble probability of the sequence $\vec{X}$, i.e. $p_{X^\dagger_{n-i:n}}(x_{i:0})=p_{X_{i:0}}(x_{i:0})$. Finally the inference strategy utilizes the time reversed dynamics $p^\dagger_{X_{n-i:n}|X_{0:n-i-1},Y_{0:n-i-1}}(\cdot,\cdot)$.
\section{Treatment of the correlated coupled Ornstein Uhlenbeck processes}
\label{appAA}
The Fokker-Planck equation in Eq.~(\ref{FP1}) has a stationary solution of the form
\begin{equation}
p^s(x,y)=(2\pi|\Psi|)^{-\frac{1}{2}}\exp{\left[\mathbf{z}^T\Psi^{-1}\mathbf{z}\right]}
\end{equation}
where 
\begin{align}
\mathbf{z}&=\left(
\begin{array}{c}
\mathbf{x}\\
\mathbf{y}\\
\end{array}
\right);\quad
\Psi=\left(
\begin{array}{cc}
\Psi_{xx}&\Psi_{xy}\\
\Psi_{yx}&\Psi_{yy}\\
\end{array}
\right)
\end{align}
and
\begin{align}
\Psi_{xx}&=\frac{(BC-AD)V_x^2+2BD\rho V_xV_y-D^2V_x^2-B^2V_y^2}{2(A+D)(AD-BC)}\nonumber\\
\Psi_{yy}&=\frac{(BC-AD)V_y^2+2CA\rho V_xV_y-A^2V_y^2-C^2V_x^2}{2(A+D)(AD-BC)}\nonumber\\
\Psi_{xy}&=\Psi_{yx}=\frac{CDV_x^2-2AD\rho V_xV_y+ABV_y^2}{2(A+D)(AD-BC)}.
\end{align}
Utilizing such a solution with Eq.~(\ref{DSint}) by first converting to It\={o} form, replacing $\langle dx\rangle$ with $Axdt+Bydt$ then averaging over $p^s(x,y)$ yields the interaction entropy production
\begin{align}
\frac{d\langle\Delta\mathcal{S}_{\rm int}^{xy}\rangle}{dt}&=\frac{\rho\left(CV_x^2-V_y((A-D)\rho V_x+BV_y)\right)\left(\rho CV_x^2-V_y((A-D) V_x+\rho BV_y)\right)}{(A+D)(\rho^2-1)V_x^2 V_y^2}
\end{align}
\section{Calculation of transfer entropy rates of the correlated coupled Ornstein Uhlenbeck processes}
\label{appBB}
In this appendix we consider the coupled stochastic differential equations
\begin{align}
dx&=Axdt+Bydt+V_xdW_x\nonumber\\
dy&=Cxdt+Dydt+V_ydW_y
\end{align}
with noise properties $\langle dW_x(t)dW_x(t')\rangle=\langle dW_y(t)dW_y(t')\rangle=\delta(t-t')dt$ and $\langle dW_x(t)dW_y(t')\rangle=\rho\delta(t-t')dt$ with $\rho\in [-1,1]$. This can be equivalently constructed by the following SDEs
\begin{align}
dx&=Axdt+Bydt+\sqrt{1-\rho^2}V_xdW_1+\rho V_xdW_2\nonumber\\
dy&=Cxdt+Dydt+V_ydW_2
\end{align}
with properties $dW_1(t)dW_2(t')=0$. This leads to a diffusion matrix
\begin{align}
\begin{bmatrix}
    \frac{V_x^2}{2}       & \frac{\rho V_xV_y}{2} \\
    \frac{\rho V_xV_y}{2}       & \frac{V_y^2}{2}
\end{bmatrix}.
\end{align}
This is a Gaussian process, since the noise is Gaussian and the driving terms are linear. Importantly, the coarse grained process in $X$ is also Gaussian.
 Critically for stationary Gaussian processes the entropy rate may be written \cite{horowitz_secondlawlike_2014}
\begin{align}
\dot{h}_x=\lim_{dt\to 0}\frac{1}{2dt}\ln(2\pi e)+\frac{1}{4\pi}\int_{-\pi/dt}^{\pi/dt}\ln \Sigma(\omega)d\omega,
\label{gaussfour}
\end{align}
such that, under certain circumstances, the transfer entropy is given by
\begin{align}
T_{y\to x}=\dot{h}_x-\dot{h}_{x|\{y\}}=-\frac{1}{4\pi}\int_{-\infty}^{\infty}\ln \frac{\Sigma_{x|\{y\}}(\omega)}{\Sigma_{x}(\omega)}d\omega
\end{align}
where $\Sigma_x(\omega)$ is the Fourier transform of the auto correlation function, $R_{xx}(\tau)=\mathbb{E}_x[x(t)x^*(t+\tau)]$, and $\Sigma_{x|\{y\}}(\omega)$ is the Fourier transform of the correlation function under the measure with knowledge of $Y$, i.e. $p[x_{t_0}^t|x_{t_0},\{y_{t_0}^t\}]$, $R_{xx|\{y\}}(\tau)=\mathbb{E}_{x|\{y\}}[x(t)x^*(t+\tau)]$. Importantly, $\Sigma_{x|\{y\}}(\omega)$ can only be computed for the $\rho=0$ case, differing from the general $\rho\neq 0$ case that we are considering.
\par
These quantities, $\Sigma_{x|\{y\}}(\omega)$ and $\Sigma_{x}(\omega)$ (with $\rho=0$ in the former) can be derived by Fourier transforming the SDEs. We write $\dot{x}=dx/dt$ and $\eta_x=V_xdW_x/dt$, such that we reformulate as Langevin equations, so that
\begin{align}
i\omega \hat{x}=A\hat{x}+B\hat{y}+\hat{\eta}_x\nonumber\\
i\omega \hat{y}=C\hat{x}+D\hat{y}+\hat{\eta}_y.
\end{align}
Since we can only identify $\Sigma_{x|\{y\}}(\omega)$ for $\rho=0$, we separate the computation of the transfer entropy rate into two components
\begin{align}
\dot{T}_{y\to x}=\dot{h}_x-\dot{h}_{x|\{y\}}=\underbrace{\dot{h}_x-\dot{h}^{\rho=0}_{x|\{y\}}}_{1}+\underbrace{\dot{h}^{\rho=0}_{x|\{y\}}-\dot{h}_{x|\{y\}}}_2.
\label{T0}
\end{align}
Firstly, we consider calculation one and so calculate $\Sigma_x(\omega)$ in the general case, $\rho \neq 0$, and $\Sigma_{x|\{y\}}(\omega)=\Sigma^{\rho=0}_{x|\{y\}}(\omega)$ in the $\rho=0$ case.
\par
To consider $\Sigma^{\rho=0}_{x|\{y\}}(\omega)$ we wish to consider the auto-correlation $\langle x'(t)x'(t+\tau)\rangle$ where $x'=x-\langle x|y\rangle$. To do so we follow \cite{horowitz_secondlawlike_2014} and consider not only the evolution of $\dot{x}$ and thus $\hat{x}$ but also $\langle \dot{x}|y\rangle$ which follows
\begin{align}
d\langle x|y\rangle&=A\langle x|y\rangle dt+Bydt\nonumber\\
i\omega\hat{\langle x|y\rangle}&=A\hat{\langle x|y\rangle}+B\hat{y}.
\end{align}
Now by the Wiener-Khinchin theorem
\begin{align}
\Sigma_{xy}(\omega)=F.T.\{R_{xy}(\tau)\}=F.T.\{\mathbb{E}[x(t)y^*(t+\tau)]\}=\mathbb{E}\left[\hat{x}\hat{y}^*\right].
\end{align}
Consequently
\begin{align}
\Sigma^{\rho=0}_{x|\{y\}}(\omega)&=F.T.\left[\langle x'(t)x'(t+\tau)\rangle\right]\nonumber\\
&=\left\langle|\hat{x}-\hat{\langle x|y\rangle}|^2\right\rangle\nonumber\\
&=\frac{\mathbb{E}\left[\hat{\eta}_x\hat{\eta}^*_x\right]}{(i\omega-A)(-i\omega-A)}\nonumber\\
&=\frac{V_x^2}{\omega^2+A^2}.
\end{align}
Solving for $\hat{x}$, independently of $y$, allows computation of $\Sigma_x(\omega)$, where
\begin{align}
\hat{x}&=\frac{(i\omega-A)(i\omega-D)}{(i\omega-A)(i\omega-D)-BC}\left(B\hat{\eta}_y+(i\omega-D)\hat{\eta}_x\right)
\end{align}
such that
\begin{align}
\Sigma_x&=\mathbb{E}[\hat{x}\hat{x}^*]\nonumber\\
&=\left|\frac{(i\omega-A)(i\omega-D)}{(i\omega-A)(i\omega-D)-BC}\right|^2\frac{1}{(A^2+\omega^2)(D^2+\omega^2)}\nonumber\\
&\qquad\times((D^2+\omega^2)\mathbb{E}[\hat{\eta}_x\hat{\eta}_x^*]+B^2\mathbb{E}[\hat{\eta}_y\hat{\eta}_y^*]-2BD(\mathbb{E}[\hat{\eta}_x\hat{\eta}_y^*]+\mathbb{E}[\hat{\eta}_x^*\hat{\eta}_y])+iB\omega(\mathbb{E}[\hat{\eta}_x\hat{\eta}_y^*]-\mathbb{E}[\hat{\eta}_x^*\hat{\eta}_y])).
\end{align}
Now similarly
\begin{align}
\mathbb{E}\left[\hat{\eta}_x\hat{\eta}^*_y\right]&=F.T.\{c(\eta_x(t),\eta_y^*(t+\tau))\}\nonumber\\
&=\frac{V_xV_ydW_x(t)dW_y(t)}{dt}\int_{-\infty}^{\infty}\delta(\tau)e^{-i\omega\tau}d\tau\nonumber\\
&=\frac{V_xV_ydW_x(t)dW_y(t)}{dt}.
\end{align}
I.e. $V_x^2$ for auto-correlation terms and $\rho V_xV_y$ for cross correlation terms. Consequently we have
\begin{align}
\mathbb{E}[\hat{\eta}_x\hat{\eta}_x^*]&=V_x^2\nonumber\\
\mathbb{E}[\hat{\eta}_y\hat{\eta}_y^*]&=V_y^2\nonumber\\
\mathbb{E}[\hat{\eta}_x\hat{\eta}_y^*]&=\mathbb{E}[\hat{\eta}_x^*\hat{\eta}_y]=\rho V_xV_y
\end{align}
and so
\begin{align}
\Sigma_x&=\left|\frac{(i\omega-A)(i\omega-D)}{(i\omega-A)(i\omega-D)-BC}\right|^2\frac{1}{(A^2+\omega^2)(D^2+\omega^2)}\nonumber\\
&\qquad\times((D^2+\omega^2)V_x^2+B^2V_y^2-2\rho BDV_x V_y).
\end{align}
Looking to the integral in Eq.~(\ref{gaussfour}) we have
\begin{align}
-\frac{1}{4\pi}\int_{-\infty}^{\infty}\ln \frac{\Sigma^{\rho=0}_{x|\{y\}}(\omega)}{\Sigma_{x}(\omega)}d\omega&=\frac{1}{4\pi}\int_{-\infty}^{\infty}\ln \left|\frac{(i\omega-A)(i\omega-D)}{(i\omega-A)(i\omega-D)-BC}\right|^2\nonumber\\
&-\frac{1}{4\pi}\int_{-\infty}^{\infty}\ln \frac{V_x^2(D^2+\omega^2)}{(D^2+\omega^2)V_x^2+B^2V_y^2-2\rho BDV_x V_y}.
\end{align}
The first of these is a known vanishing integral \cite{horowitz_secondlawlike_2014}. For the second we exploit the integral
\begin{align}
\int_0^\infty\ln\frac{x^2+a^2}{x^2+b^2}dx=\pi(a-b)
\end{align}
to arrive at
\begin{align}
\dot{h}_{x}-\dot{h}^{\rho=0}_{x|\{y\}}&=\frac{|D|}{2}\left(\sqrt{1+\frac{BV_y}{DV_x}\left(\frac{BV_y}{DV_x}-2\rho\right)}-1\right),
\label{T1}
\end{align}
but this is only half of the solution. Next we need to consider $\dot{h}^{\rho=0}_{x|\{y\}}-\dot{h}_{x|\{y\}}$. To do this we consider a different approach. 
\par
Since the system is linear/Gaussian, the entropy rate is entirely encoded by the variance in the $\mathcal(dt)$ regime. That is we have
\begin{align}
\dot{h}dt=\frac{1}{2}\ln(2\pi e \sigma^2dt)+\mathcal{O}(dt^2).
\label{hrate}
\end{align}
To calculate this we explicitly describe an expression for  the solution to the original Langevin equations (which, incidentally, is of closed form when $C=0$). We do this individually for both $X$ and $Y$ by writing
\begin{align}
\frac{d}{dt}\left(e^{-At}x(t)\right)&=\left(By(t)+\eta_x(t)\right)e^{-At}\nonumber\\
\frac{d}{dt}\left(e^{-Dt}y(t)\right)&=\left(Cx(t)+\eta_y(t)\right)e^{-Dt}
\end{align}
such that
\begin{align}
x(t)&=x(0)e^{At}+\int_0^te^{A(t-t')}By(t')dt'+\int_0^te^{A(t-t')}\eta_x(t')dt'\nonumber\\
y(t)&=y(0)e^{Dt}+\int_0^te^{D(t-t')}Cx(t')dt'+\int_0^te^{D(t-t')}\eta_y(t')dt'.
\end{align}
We then substitute $y(t)$ in for $y(t')$ such that
\begin{align}
x(t)&=x(0)e^{At}+\int_0^te^{A(t-t')}B\left[y(0)e^{Dt'}+\int_0^{t'}e^{D(t'-t'')}Cx(t'')dt''+\int_0^{t'}e^{D(t'-t'')}\eta_y(t'')dt''\right]dt'+\int_0^te^{A(t-t')}\eta_x(t')dt'.
\end{align}
When $C\neq 0$ it is not of closed form, but adequate for our purposes. From this we want to find $\langle(x(t)-\langle x(t)\rangle)^2\rangle$ where the average $\langle x\rangle$ is shorthand for $\langle x(t)|x(0),y(0)\rangle$. We achieve this by noting
\begin{align}
\left\langle\int_0^t f(t')\eta(t')dt'\right\rangle=\int_0^t f(t')\left\langle\eta(t')\right\rangle dt'=0
\end{align}
since we have $\left\langle\eta(t')\right\rangle=0$. As such we have
\begin{align}
x(t)-\langle x(t)\rangle&=B\int_0^te^{A(t-t')}dt'\int_0^{t'}dt''e^{D(t'-t'')}\eta_y(t'')+\int_0^tdt'e^{A(t-t')}\eta_x(t')
\end{align}
safely independent of $C$. Consequently the variance is then given by
\begin{align}
\langle(x(t)-\langle x(t)\rangle)^2\rangle=&\left\langle\int_0^tdt'\int_0^tdt''e^{A(2t-t'-t'')}\eta_x(t')\eta_x(t'')\right\rangle\nonumber\\
&+2B\left\langle\int_0^tdt'\int_0^{t'}dt''\int_0^tdt'''e^{A(2t-t'-t''')+D(t'-t'')}\eta_x(t''')\eta_y(t'')\right\rangle\nonumber\\
&+B^2\left\langle\int_0^tdt'\int_0^{t'}dt''\int_0^tdt'''\int_0^{t'''}dt''''e^{A(2t-t'-t''')+D(t'+t'''-t''-t'''')}\eta_y(t'')\eta_y(t'''')\right\rangle\nonumber\\
=&\int_0^tdt'\int_0^tdt''e^{A(2t-t'-t'')}\left\langle\eta_x(t')\eta_x(t'')\right\rangle\nonumber\\
&+2B\int_0^tdt'\int_0^{t'}dt''\int_0^tdt'''e^{A(2t-t'-t''')+D(t'-t'')}\left\langle\eta_x(t''')\eta_y(t'')\right\rangle\nonumber\\
&+B^2\int_0^tdt'\int_0^{t'}dt''\int_0^tdt'''\int_0^{t'''}dt''''e^{A(2t-t'-t''')+D(t'+t'''-t''-t'''')}\left\langle\eta_y(t'')\eta_y(t'''')\right\rangle.
\end{align}
Now we identify
\begin{align}
\left\langle\eta_x(t')\eta_x(t'')\right\rangle&=\frac{V_x^2}{2}\delta(t''-t')\nonumber\\
\left\langle\eta_y(t')\eta_y(t'')\right\rangle&=\frac{V_y^2}{2}\delta(t''-t')\nonumber\\
\left\langle\eta_x(t')\eta_y(t'')\right\rangle&=\frac{\rho V_xV_y}{2}\delta(t''-t')
\end{align}
and also the usual sifting property
\begin{align}
\int_0^t\left[\int_0^{t'}f(t''',t'')\delta(t'''-t'')dt'''\right]dt''&=\int_0^t1_{[0,t']}(t'')f(t'',t'')dt''\nonumber\\
&=\int_0^{\min(t,t')}f(t'',t'')dt''\nonumber\\
&=\int_0^{t'}f(t'',t'')dt''\qquad (t>t').
\end{align}
Consequently,
\begin{align}
\langle(x(t)-\langle x(t)\rangle)^2\rangle=&\frac{V_x^2}{2}\int_0^tdt'e^{2A(t-t')}\nonumber\\
&+\rho BV_xV_y\int_0^tdt'\int_0^{t'}dt'''e^{A(2t-t't''')+D(t'-t''')}\nonumber\\
&+B^2\frac{V_y^2}{2}\int_0^tdt'\int_0^tdt'''\int_0^{\min(t',t''')}dt''e^{A(2t-t'-t''')+D(t'+t'''-2t'')}\nonumber\\
=&\frac{V_x^2}{2}\frac{e^{2At}-1}{2A}\nonumber\\
&+\rho BV_xV_y\frac{A-D+(A+D)e^{2At}-2Ae^{(A+D)t}}{2(A^3-AD^2)}\nonumber\\
&+B^2\frac{V_y^2}{2}\left[\int_0^tdt'\int_{t'}^t dt'''\int_0^{t'}dt''+\int_0^tdt'''\int_{t'''}^t dt'\int_0^{t'''}dt''\right]e^{A(2t-t'-t''')+D(t'+t'''-2t'')}\nonumber\\
=&\frac{V_x^2}{2}\frac{e^{2At}-1}{2A}\nonumber\\
&+\rho BV_xV_y\frac{A-D+(A+D)e^{2At}-2Ae^{(A+D)t}}{2(A^3-AD^2)}\nonumber\\
&+B^2\frac{V_y^2}{2}\frac{D^2(e^{2At}-1)+A^2(e^{2Dt}-1)+AD\left(2+e^{2At}+e^{2Dt}-4e^{(A+D)t}\right)}{2AD(A+D)(A-D)^2}.
\end{align}
The first term is $\mathcal{O}(t)$, the second $\mathcal{O}(t^2)$ and the third $\mathcal{O}(t^3)$. Examining the form of Eq.~(\ref{hrate}) we need second order contributions in $\sigma^2$ in order to capture first order terms in $\dot{h}dt$ and so we identify, expanding the exponentials,
\begin{align}
\langle(x(t)-\langle x(t)\rangle)^2\rangle=&\frac{V_x^2t}{2}\left(1+At+\rho B\frac{V_y}{V_x}t\right)+\mathcal{O}(t^3)
\end{align}
such that
\begin{align}
\dot{h}_{x|\{y\}}dt=\frac{1}{2}\ln{[\pi e V_x^2dt]}+\frac{A}{2}dt+\frac{\rho B V_y}{2V_x}dt
\end{align}
and therefore
\begin{align}
\dot{h}^{\rho=0}_{x|\{y\}}-\dot{h}_{x|\{y\}}=-\frac{\rho B V_y}{2V_x}.
\end{align}
Putting this together with Eq.~(\ref{T0}) and Eq.~(\ref{T1}) we finally get the result
\begin{align}
\dot{T}_{y\to x}&=\frac{|D|}{2}\left(\sqrt{1+\frac{BV_y}{DV_x}\left(\frac{BV_y}{DV_x}-2\rho\right)}-\left(1+\rho\frac{BV_y}{|D|V_x}\right)\right)
\label{Tfin}
\end{align}
with the appropriate $\dot{T}_{x\to y}$ result found by simply exchanging variables $V_x\leftrightarrow V_y$, $A\leftrightarrow D$, $B\leftrightarrow C$, due to symmetry. We note the lack of dependence on $A$ and $C$. 
\end{widetext}
\end{appendix}
\bibliographystyle{apsrev4-1}
\bibliography{ais2_arxiv2}
\end{document}